\documentclass[aip,rsi,reprint,graphicx]{revtex4-1} 

\usepackage{amsmath}    
\usepackage{verbatim}   
\usepackage{color}      
\usepackage{subfigure}  
\usepackage{placeins}	
\usepackage[pdftex]{graphicx} 
\usepackage{epstopdf}
\usepackage{hyperref}
\hypersetup{
  colorlinks=true,
  linkcolor=blue,
  citecolor=blue,
  urlcolor=blue,
  breaklinks=true}
 
\usepackage{units}   
\usepackage[separate-uncertainty=true]{siunitx}    
  
 \newcommand{\alca}{$^{27}\mathrm{Al}^+/\,^{40}\mathrm{Ca}^+$ }

 \newcommand{\sods}{$\frac{\delta\nu_\mathrm{EMM}}{\nu}=\left({-0.4}^{+{0.4}}_{-{0.3}}\right)\times10^{-18}$ } 
 
 \newcommand{\al}{$^{27}\mathrm{Al}^+$}
 \newcommand{\ca}{$^{40}\mathrm{Ca}^+$ }
 \newcommand{\oaxca}{\omega_\mathrm{ax,Ca+} } 
 \newcommand{\oradca}{\omega_\mathrm{rad,Ca+} } 

\begin{document}

\title{Towards a Transportable Aluminium Ion Quantum Logic Optical Clock} 

\author{S. Hannig}
\affiliation{Physikalisch-Technische Bundesanstalt, Bundesallee 100, 38116 Braunschweig, Germany}

\author{L. Pelzer}
\affiliation{Physikalisch-Technische Bundesanstalt, Bundesallee 100, 38116 Braunschweig, Germany}

\author{N. Scharnhorst}
\affiliation{Physikalisch-Technische Bundesanstalt, Bundesallee 100, 38116 Braunschweig, Germany}
\affiliation{Institut f{\"u}r Quantenoptik, Leibniz Universit{\"a}t Hannover, Welfengarten 1, 30167 Hannover, Germany}

\author{J. Kramer}
\affiliation{Physikalisch-Technische Bundesanstalt, Bundesallee 100, 38116 Braunschweig, Germany}

\author{M. Stepanova}
\affiliation{Physikalisch-Technische Bundesanstalt, Bundesallee 100, 38116 Braunschweig, Germany}
\affiliation{Institut f{\"u}r Quantenoptik, Leibniz Universit{\"a}t Hannover, Welfengarten 1, 30167 Hannover, Germany}

\author{Z. T. Xu}
\affiliation{MOE Key Laboratory of Fundamental Physical Quantities Measurement, School of Physics, Huazhong University of Science and Technology, 430074  Wuhan, P. R. China}

\author{N. Spethmann}
\affiliation{Physikalisch-Technische Bundesanstalt, Bundesallee 100, 38116 Braunschweig, Germany}

\author{I. D. Leroux}
\altaffiliation{Current address: National Research Council Canada, Ottawa, Ontario, K1A 0R6, Canada}
\affiliation{Physikalisch-Technische Bundesanstalt, Bundesallee 100, 38116 Braunschweig, Germany}

\author{T. E. Mehlst{\"a}ubler}
\affiliation{Physikalisch-Technische Bundesanstalt, Bundesallee 100, 38116 Braunschweig, Germany}

\author{P. O. Schmidt}
\email[Corresponding author: ]{Piet.Schmidt@ptb.de}
\affiliation{Physikalisch-Technische Bundesanstalt, Bundesallee 100, 38116 Braunschweig, Germany}
\affiliation{Institut f{\"u}r Quantenoptik, Leibniz Universit{\"a}t Hannover, Welfengarten 1, 30167 Hannover, Germany}

\date{\today}

\begin{abstract}
With the advent of optical clocks featuring fractional frequency uncertainties on the order of  $10^{-17}$ and below, new applications such as chronometric levelling with few-cm height resolution emerge. We are developing a transportable optical clock based on a single trapped aluminium ion, which is interrogated via quantum logic spectroscopy. We employ singly-charged calcium as the logic ion for sympathetic cooling, state preparation and readout. Here we present a simple and compact physics and laser package for manipulation of $^{40}\mathrm{Ca}^+$. Important features are a segmented multi-layer trap with separate loading and probing zones, a compact titanium vacuum chamber, a near-diffraction-limited imaging system with high numerical aperture based on a single biaspheric lens, and an all-in-fiber \ca repump laser system. We present preliminary estimates of the trap-induced frequency shifts on \al, derived from measurements with a single calcium ion.
The micromotion-induced second-order Doppler shift for \al has been determined to be \sods and the black-body radiation shift is $\delta\nu_\mathrm{BBR}/\nu=(-4.0\pm0.4)\times10^{-18}$. Moreover, heating rates of 30 (7) quanta per second at trap frequencies of $\oradca\approx2\pi\times2.5\,\mathrm{MHz}$ ($\oaxca\approx2\pi\times1.5\,\mathrm{MHz}$) in radial (axial) direction have been measured, enabling interrogation times of a few hundreds of milliseconds.
\end{abstract}


\maketitle

\section{Introduction}
\label{sec:intro}
For more than 50~years the definition of the SI-second has been based on the transition frequency between the two hyperfine levels of the ground state of the caesium-133 atom, which lies in the microwave regime. The SI-second is realized in caesium fountain primary frequency standards, achieving a fractional frequency inaccuracy in the low $10^{-16}$ range\cite{guena_first_2017,heavner_first_2014,abgrall_atomic_2015,weyers_advances_2018}. 
Clocks based on optical transitions facilitate substantially lower fractional frequency uncertainties, as their transition frequencies can be up to $10^5$ times higher for comparable linewidths. In the last decade, optical clocks\cite{ludlow_optical_2015,poli_optical_2013} surpassed the best microwave fountain clocks by achieving $10^{-17}$ estimated systematic uncertainties \cite{huntemann_single-ion_2016, chou_frequency_2010, nicholson_systematic_2015, ushijima_cryogenic_2015,mcgrew_atomic_2018}. Besides enabling a possible future redefinition of the SI-second\cite{gill_when_2011,riehle_towards_2015}, they also allow for the search for a variation of fundamental constants \cite{godun_frequency_2014,huntemann_improved_2014,safronova_search_2017}, and pave the way towards new applications such as chronometric leveling \cite{bjerhammar_relativistic_1985, vermeer_chronometric_1983}, where the differential gravitational red shift is measured by comparing two spatially separated clocks to determine their height difference \cite{muller_high_2018, mehlstaubler_atomic_2018}. Transportable optical clocks that have been calibrated to a reference clock are required to facilitate this novel approach by deploying them at geodetically relevant sites and comparing them to the reference clock e.g.~ through phase-stabilized optical fibre links \cite{lisdat_clock_2016,raupach_brillouin_2015, riehle_optical_2017, lee_hybrid_2017}. A transportable optical single $\mathrm{Ca}^+$ clock has been reported\cite{cao_transportable_2017} to reach a systematic uncertainty of $7.8\times10^{-17}$ in a system occupying a volume of less than $1\,\mathrm{m}^3$. Furthermore, a $^{87}\mathrm{Sr}$ optical lattice clock with an estimated fractional systematic uncertainty of $7\times 10^{-17}$ has been installed in a car trailer\cite{koller_transportable_2017}. It has been employed in the first measurement campaign involving a transportable optical clock to determine the gravitational red shift of the Modane underground laboratory relative to clocks at INRIM (Turin)\cite{grotti_geodesy_2018}. 

Aluminium is a promising candidate for a high-accuracy single-ion clock, since it has the smallest black-body radiation shift of all realised optical frequency references \cite{rosenband_blackbody_2006, safronova_precision_2011}. Therefore, the difficult-to-control temperature environment seen by the atoms needs to be known to only 18.2~K at room temperature \cite{dolezal_analysis_2015} to reach $10^{-18}$ systematic uncertainty of this shift, instead of the few-10~mK uncertainty required for e.g. Sr \cite{nicholson_systematic_2015}. Moreover, $\mathrm{Al}^+$ has only small linear and quadratic Zeeman shifts\cite{rosenband_observation_2007} and negiglible quadrupole shift\cite{beloy_hyperfine-mediated_2017}. The dominant shifts for $\mathrm{Al}^+$ are second-order Doppler shifts from secular and micromotion, which are both trap related and will be addressed in this work.

Direct cooling and state detection of $\mathrm{Al}^+$ requires deep ultra violet (UV) radiation at 167~nm, which to our knowledge is not commercially available. This challenge can be overcome by quantum logic spectroscopy\cite{schmidt_spectroscopy_2005}, where a clock ion and an additional so-called logic ion of a different species are confined in the same trap. The logic ion provides sympathetic cooling to the clock ion in case of the $\mathrm{Al}^+$ clock\cite{wubbena_sympathetic_2012}. For readout, the internal state of the clock ion is transferred to the logic ion via a shared motional mode using a series of laser pulses. Afterwards, the internal state of the logic ion is determined by electron shelving detection.

The first $\mathrm{Al}^+$ clocks were operated at NIST\cite{rosenband_observation_2007,rosenband_frequency_2008, chou_frequency_2010}, and have been evaluated to a fractional systematic uncertainty of as low as $8.6\times10^{-18}$. Using these clocks, the first optical-clock-based laboratory measurement of the dependence of the gravitational red shift on a change in the height difference has been performed\cite{chou_optical_2010}. Since then, a number of groups have started new $\mathrm{Al}^+$ clock setups\cite{guggemos_precision_2016,zhang_direct_2017,cui_sympathetic_2018}, including our own at PTB \cite{wubbena_controlling_2014,scharnhorst_experimental_2018}.

Here we report on the characterization of PTB's second Al ion clock setup which is operated with \ca as logic ion. It is designed for transportability to enable side-by-side comparisons with other clocks and to perform chronometric levelling in a non-laboratory environment. The entire physics package is non-magnetic, since all metal parts are made of titanium. The measurements are conducted in a laser-cut segmented multi-layer ion trap made of the printed circuit board material Rogers\textsuperscript{\textregistered}~4350. The trap includes on-chip filtered compensation electrodes and temperature sensors\cite{herschbach_linear_2012,pyka_high-precision_2014}. Atomic beams of the required species are produced by ablation\cite{hendricks_all-optical_2007,guggemos_sympathetic_2015} with a pulsed Nd:YAG laser that is frequency doubled to \SI{532}{\nm}. A photoionization pulse, following the ablation pulse after an adjustable time of flight (TOF), allows kinetic-energy-selective loading of \al.
All required wavelengths for $\mathrm{Ca}^+$ cooling, repumping and coherent manipulation are generated directly from diode lasers. The repumper laser system for $\mathrm{Ca}^+$ is fully fiberized and therefore compact and robust. All laser systems are mounted on breadboards that can be stacked up to reduce the footprint of the system.

We demonstrate single $\mathrm{Ca}^+$ imaging with a diffraction-limited resolution of \SI{0.79}{\um} in the object plane by using a custom biaspheric lens and a compact standard scientific complementary metal-oxide-semiconductor (CMOS) camera. \SI{10}{\percent} of the fluorescent light is directed to the camera, while \SI{90}{\percent} is directed towards a photomultiplier tube (PMT). With the CMOS camera we obtain a signal-to-noise ratio (SNR) of 80 for a region of interest around a single ion and a \SI{300}{\ms} exposure time, while the PMT provides state discrimination with an error of below \SI{0.2}{\percent} for \SI{25}{\us} detection time.
We have compensated excess micromotion down to an overall residual ac electric field of $(67.8\pm6.9)\,\mathrm{V/m}$ at the trap center, which corresponds to a fractional frequency second-order Doppler shift of \sods for a single $\mathrm{Al}^+$.
Ground state cooling of a single $\mathrm{Ca}^+$ has been achieved in all three modes.

The paper is structured as follows: in Section~\ref{sec:ions} we introduce the reduced term schemes of the involved ions. In Section~\ref{sec:setup} we describe the experimental setup. The results on the imaging performance, micromotion compensation, heating rate measurements, and $\mathrm{Al}^+$ initial cooling times after ablation loading are presented in Section~\ref{sec:experiment}, including a partial uncertainty budget.

\section{Ion species}
\label{sec:ions}
\begin{figure}[htb!]
\includegraphics[width=8.0cm]{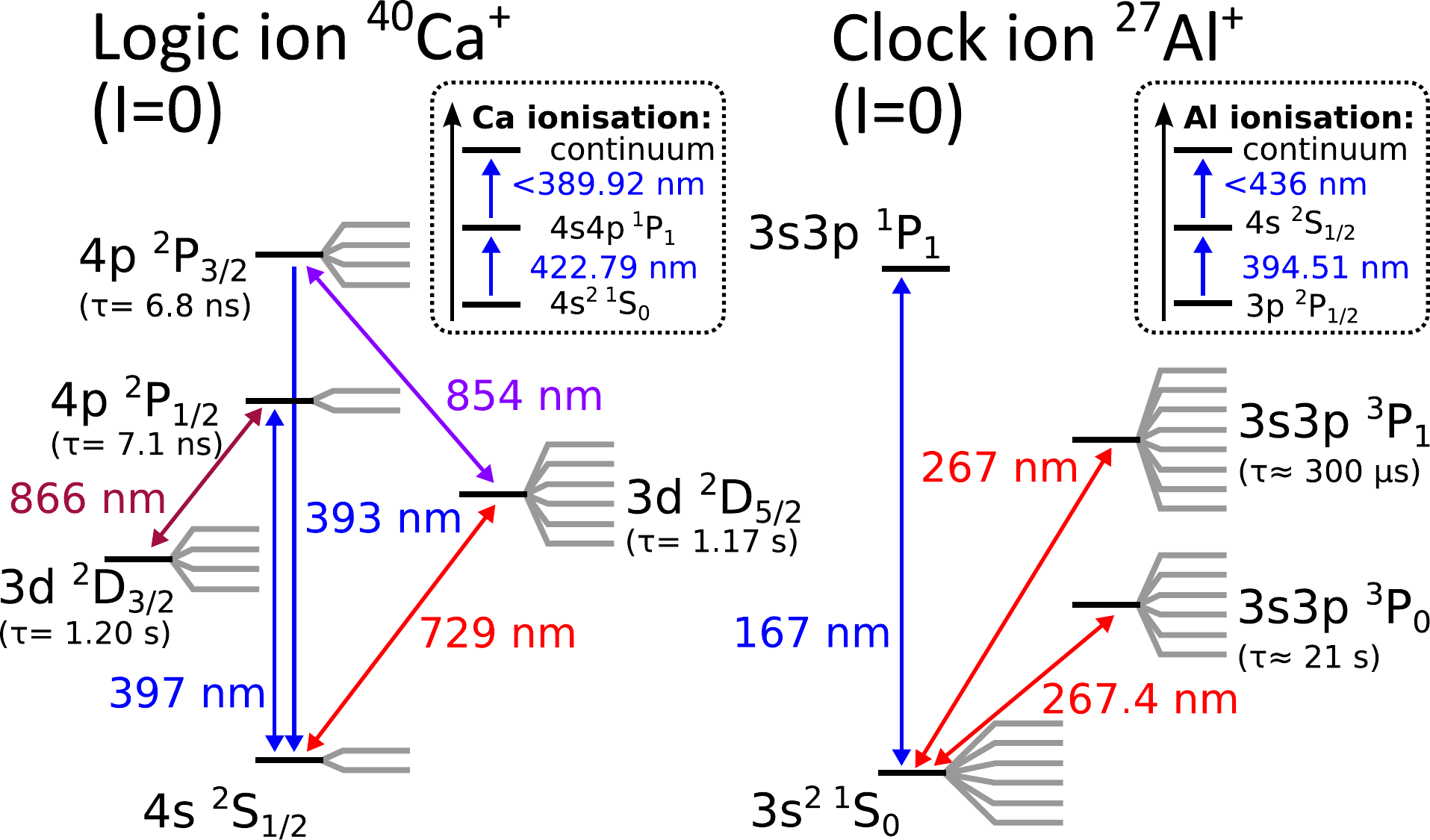}
\caption{Reduced $\mathrm{Ca}^+$ and $\mathrm{Al}^+$ level schemes (not to scale). Grey lines depict the Zeeman levels when an external magnetic field is applied. The insets show the employed ionization schemes for neutral atoms of both species consisting of a first resonant and a second non-resonant step in both cases.}
\label{fig:levels}
\end{figure}

Fig.~\ref{fig:levels} shows the reduced level schemes for \al and \ca including the transitions relevant for a \alca clock. Aluminium features an \SI{8}{\mHz} narrow clock transition  at \SI{267.4}{\nm} connecting the $^1$S$_0$ ground state to the $(20.6\pm1.4)\,\mathrm{s}$ long lived\cite{rosenband_observation_2007} $^3$P$_0$ level and a ``logic'' transition at \SI{266.9}{\nm}\cite{rosenband_+_2005} connecting the $^1$S$_0$ ground state to the $^3$P$_1$, $F=7/2$ excited state suitable for state preparation and readout. 
We employ sympathetic cooling\cite{larson_sympathetic_1986,barrett_sympathetic_2003} of the clock ion with an ion of another species and quantum logic spectroscopy\cite{schmidt_spectroscopy_2005, rosenband_frequency_2008} to transfer the internal state of \al to the logic ion via a shared motional mode. The logic ion needs to provide accessible closed transitions for Doppler cooling and electron shelving detection\cite{dehmelt_proposed_1973}. Furthermore, a mass ratio of the species close to unity is advantageous to enable efficient sympathetic cooling\cite{wubbena_sympathetic_2012}.
\ca is an advantageous choice as logic ion especially for a transportable clock, since all required wavelengths can be generated in compact and reliable diode lasers and guided in commercially available optical fibers, which allow for a modular setup. Moreover, quantum logic operations on \ca are very well developed through its use in quantum information processing\cite{haffner_quantum_2008}, while its mass still matches well with \al. 

\ca provides a cooling transition at 396.847~nm \cite{wan_precision_2014} with a linewidth of 22.4~MHz, inferred from its 7.1\,ns\cite{jin_precision_1993} upper state lifetime. Since the upper $^2$P$_{1/2}$ level decays with a branching ratio of $0.06435(7)$ \cite{ramm_precision_2013} to the metastable $^2$D$_{3/2}$ level with a lifetime\cite{shao_direct_2018} of 1.195(8)\,s, a repumping laser at \SI{866.214}{\nm}\cite{gebert_precision_2015} is required. Electron shelving is implemented on the 137~mHz narrow quadrupole transition\cite{shao_direct_2018} at \SI{729.147}{\nm}\cite{chwalla_absolute_2009, huang_frequency_2016}. For initialisation and sideband cooling, the upper $^2$D$_{5/2}$ level is cleared out via a \SI{854.209}{\nm} laser\cite{NISTDATA} to the \SI{6.8}{\ns} short lived\cite{NISTDATA} $^2$P$_{3/2}$ level which decays predominantly to the ground state, quenching the \SI{729}{\nm} transition.

\section{Experimental setup}
\label{sec:setup}

\begin{figure}[htb!]
\includegraphics[width=8.0cm]{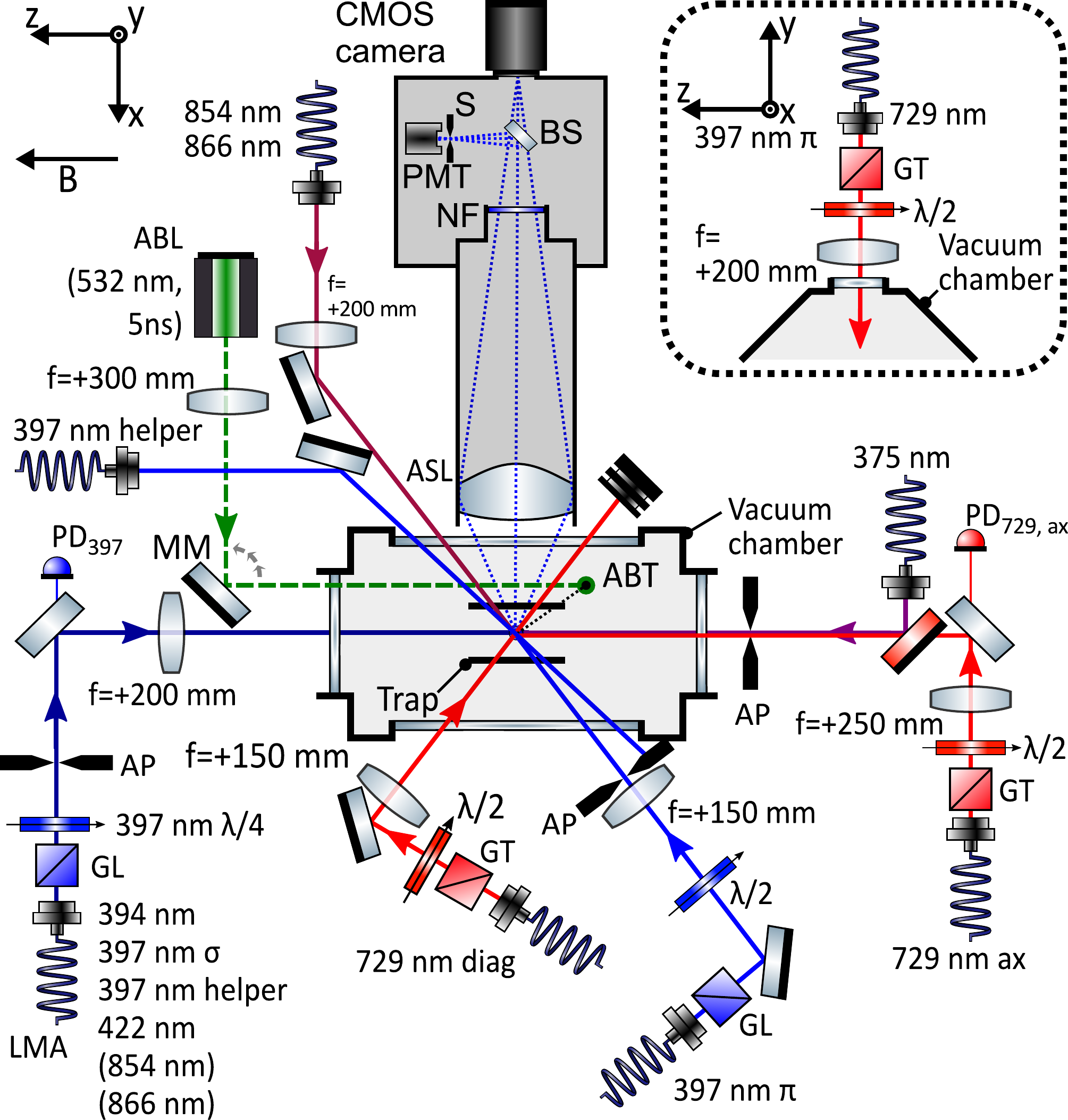}
\caption{Schematic horizontal cross section of the vacuum chamber and the surrounding optics (not to scale). The ablation laser (ABL) spot on the ablation target (ABT) is aligned via the motorized mirror (MM). Fluorescence light is collected with a 42.6~mm effective diameter biaspheric lens (ASL) through a viewport. Behind a spectral notch filter (NF), a beamsplitter (BS) directs \SI{10}{\percent} on a CMOS camera and 90\% through a spatial filter onto a photo-multiplier tube (PMT). The imaging system has been designed for 18-fold magnification. All beams required for $\mathrm{Ca}^+$ loading and coherent manipulation are delivered via various polarization-maintaining and large mode area (LMA) glass fibers. GL: Glan Laser polarizer, GT: Glan Taylor polarizer, $\lambda/n$: waveplate, AP: aperture, PD$_i$: power monitoring, lenses are specified by their focal length $f$. Insert: Vertical cross section with optics on top viewport. Three orthogonal pairs of magnetic field coils surrounding the chamber are not depicted for clarity.}
\label{fig:achamber}
\end{figure}

Fig.~\ref{fig:achamber} shows two schematic cross sections of the vacuum chamber and the surrounding optics. We chose a segmented multi-layer linear Paul trap with an ion-electrode distance of \SI{700}{\um} based on precisely aligned stacked \SI{250}{\um} thick Rogers\textsuperscript{TM} printed circuit boards as described by Herschenbach \textit{et al.}\cite{herschbach_linear_2012} for its low excess micromotion \cite{pyka_high-precision_2014} and heating rates \cite{keller_spectroscopic_2015}. The electrode structures are laser cut. All conductive structures are coated with a NiPdAu plating, where a \SIrange{3}{7}{\um} thick nickel layer provides adhesion for a \SIrange{0.05}{0.25}{\um} thick palladium and \SI{300}{\nm} thick gold layer. The stack is glued onto a rigid carrier board made of AlN that provides a good thermal connection to the environment. The trap features multiple zones for loading and operating the clock. The experiment zones are 1.0, 1.5, and \SI{2.0}{\mm} long, respectively. These different lengths allow for a compromise between the maximum achievable axial trap frequency for a given voltage and the homogeneity of the axial field\cite{herschbach_linear_2012}. Moreover, it is possible to e.g.\ trap multiple ion ensembles in different zones of the trap simultaneously, allowing for dead time-free interleaved interrogation of two clocks\cite{schioppo_ultrastable_2017} in a single trap or other multi-ensemble protocols\cite{rosenband_exponential_2013,borregaard_efficient_2013} in the future.

Several small mirrors are mounted directly on the trap in preparation for future interferometric stabilization of the path length from the clock laser to the trap. Radial ion confinement is provided by a radio-frequency (rf) potential, driven by a frequency generator (Marconi 2024; all brand names are given for illustrative purpose and not a sign of endorsement) operating at \SI{24.65}{\MHz}. This is the "magic" drive frequency for $\mathrm{Ca}^+$, at which the second-order Doppler and ac Stark shift of the trap drive cancel\cite{berkeland_minimization_1998,huang_frequency_2016}. The rf voltage is amplified and the drive circuit impedance matched to the trap by a helical resonator\cite{macalpine_coaxial_1959, siverns2012} with a loaded quality factor of $Q\approx300$. Low-noise and finely-tunable dc voltages\cite{beev_low-drift_2017} are applied to the trap electrodes for axial confinement and radial micromotion compensation \cite{berkeland_minimization_1998, pyka_high-precision_2014,keller_precise_2015}. All dc electrode voltages are filtered directly on the trap boards by first-order RC low-pass filters with a cutoff frequency of  $\sim100$~Hz. The vacuum chamber is surrounded by three pairs of magnetic field coils (not shown in the figure) to compensate for external magnetic fields and to define the quantization field $B_z=\SI{200}{\micro\tesla},$ along the trap axis.

The trap is mounted in an octagonal vacuum vessel (Kimball Physics MCF800-SphOct-G2C8) made of titanium alloy (Ti-6Al-4V) with anti-reflective (AR) coated UV fused silica vacuum windows mounted in titanium flanges. The vacuum is maintained by a compact non-evaporative getter and ion pump (SAES Getters NEXTORR D 200-5) and measured using a standard hot cathode gauge. Several broadband AR-coated CF40 and two CF160 viewports grant adequate optical access to the trap volume. Since the latter ones protrude beyond the knife-edge into the vacuum by about \SI{25}{\mm}, the biaspheric lens (ASL) optimized for diffraction-limited imaging of the \SI{397}{\nm} fluorescence photons can be as little as \SI{36.6}{\mm} from the ion. With its effective diameter of \SI{42.6}{\mm} it covers about \SI{7}{\percent} solid angle. The fluorescence light is spectrally filtered by a notch filter (NF, Thorlabs FBH400-40) and split 90:10 between a photo-multiplier tube (PMT, Hamamatsu H10682) and a sCMOS camera (PCO edge 4.2LT). While the image plane of the ASL coincides with the CMOS chip, an aperture for spatial filtering is placed in the focus in front of the PMT. In order to allow for a convenient change of the observed trap zone, the entire enclosed imaging system is mounted on a motorized 1d-translation stage to be aligned parallel to the trap axis.

The laser beams for photoionization, cooling and coherent manipulation of $^{40}\mathrm{Ca}^{+}$ are delivered through various single-mode polarization-maintaining optical fibers and then focused down to spot diameters between 60 and \SI{180}{\um} at the trap center. In the case of the \SI{397}{\nm} and \SI{729}{\nm} beams, the polarization is cleaned by Glan Laser (GL) and Glan Taylor (GT) polarizers, respectively, and then set by half or quarter waveplates ($\lambda/n$). Various photodiodes ($\mathrm{PD}_{i}$) placed behind backside-polished mirrors allow for optical power monitoring and could be used for power stabilisation.

\begin{figure}[htb!]
\includegraphics[width=8.0cm]{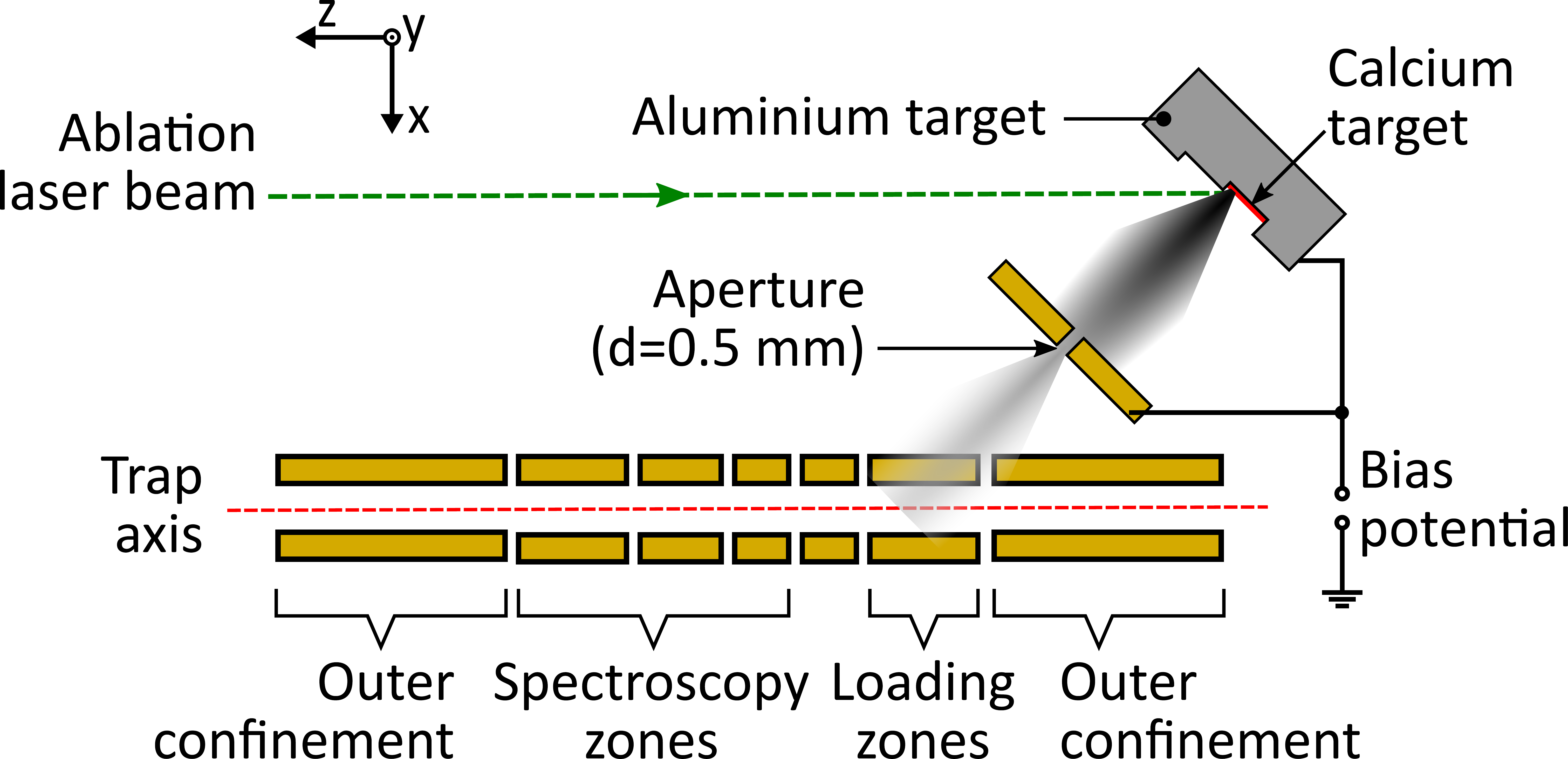}
\caption{Schematic horizontal cut through the trap center and the ablation target. The ablation laser beam can be tilted between the different targets to select either species to be loaded. An aperture restricts contamination with ablated material to the loading zone. The target-to-trap center distance is \SI{17}{\mm}. All photoionization lasers are aligned along the trap axis.}
\label{fig:trabl}
\end{figure}

A Q-switched and frequency doubled Nd:YAG laser (ABL) at \SI{1064}{\nm} (Continuum Minilite 1) is used for neutral atom beam generation via pulsed ablation from solid targets\cite{Dreyfus_1986, Towrie_1990, hendricks_all-optical_2007, zimmermann_laser_2012, guggemos_sympathetic_2015}. The beam is focused down to a spot diameter of $\approx\SI{200}{\um}$ and can be tilted in two dimensions using a motorized mirror (MM) for target selection. Fig.~\ref{fig:trabl} shows a rendering of the ablation target and the trap cut in the horizontal plane. A solid structure made of aluminium acts as Al ablation target. Calcium grains were glued (Thorlabs 353NDPK Epoxy) in a cutout in the structure (red zone). An aperture between the ablation targets and the trap shapes the plume of neutral atoms (black) generated during ablation into a collimated beam. Additionally, a negative bias voltage can be applied on the target to prevent unwanted ions generated by the ablation pulse from entering the trap. After loading the desired ion crystal in the loading zone, it can be shuttled to one of the three spectroscopy zones discussed above.

\begin{figure}[htb!]
\includegraphics[width=8.0cm]{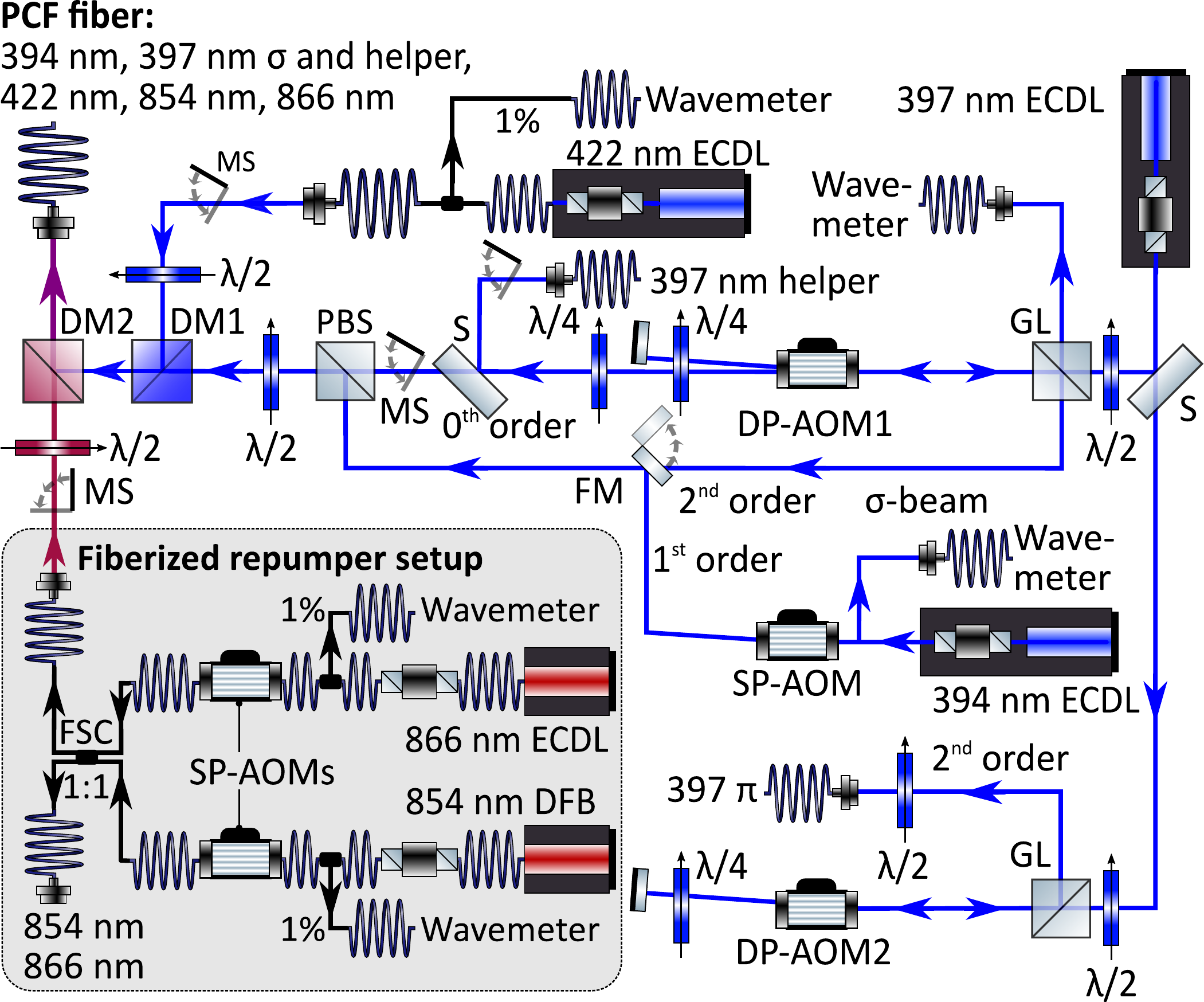}
\caption{Schematic layout of the $\mathrm{Ca}^+$ cooling laser system including the two resonant photoionization lasers for both species. The \ca cooling laser at \SI{397}{\nm} is split into two paths through a double-pass (DP) AOM each, generating independent beams for $\sigma$- and $\pi$-polarization (see also Fig.~\ref{fig:achamber}). The two repumper lasers are part of a compact fully fiberized setup, including optical isolators, fiber-taps to the wavelength meter, fiberized single-pass (SP) AOMs and fiber-combiners (FSC). Both beams are overlapped with the resonant photoionization laser beams at \SI{422}{\nm} for Ca and \SI{394}{\nm} for Al, before they are coupled into a large mode area photonic crystal fiber (PCF) to be delivered on the trap axis.
DM: Dichroic mirror, FM: Flip-mirror, GL: Glan Laser, MS: motorized shutter, PBS: polarizing beam splitter, S: 50/50 splitter, $\lambda/n$: waveplate.}
\label{fig:ca-lasers}
\end{figure}

Fig.~\ref{fig:ca-lasers} shows a schematic overview of the most important parts of the laser system. Laser cooling is implemented with an external diode laser (ECDL\footnote{All employed ECDLs are Toptica DL Pro Systems.}) at \SI{397}{\nm} that is split into two paths. Each path contains a double-pass acousto-optical modulator (AOM) for frequency tuning by $\pm50\,\mathrm{MHz}$. The zeroth order of DP-AOM1 is split into two far detuned helper beams for rapid cooling of hot ions\cite{wubbena_sympathetic_2012}. One of the helper beams is overlapped with the first order of the same AOM after double passing, the resonant Ca photoionization laser at \SI{422}{\nm}, and the two repumper beams, and then coupled into a large mode area (LMA) photonic crystal fiber (NKT LMA-PM-5) to be delivered along the trap axis as shown in Fig.~\ref{fig:achamber}. The second helper is applied diagonally through the loading zone and provides 3d laser cooling. Both helper beams are shuttered individually by mechanical shutters (MS). This configuration allows for loading and cooling in the loading zone, and shuttling the ions to the spectroscopy zone without any laser realignments. The repumper beams are generated in a fully fiberized part of the setup, starting with an \SI{866}{\nm} ECDL and \SI{854}{\nm} DFB (Eagleyard photonics EYP-DFB-0853-00050-1500-BFY02-0000) laser, which are sent through individual fiberized optical isolators. In two fiber splitters \SI{1}{\percent} of the power is coupled to a wavelength meter (WLM). Two fiber-coupled single-pass AOMs are used as fast shutters and for fine-tuning of the laser frequency with a 3~dB bandwidth of $\pm 15$~MHz. Behind those, both wavelengths enter a $50\%$~fiber combiner, which splits equal contributions to both of its output ports. To reduce power fluctuations after the polarization-cleaning elements at the fiber exits, all the aforementioned fibers are polarization-maintaining.

A fraction of the light of each laser is sent via two single-mode fiber switches to a wavelength meter (High Finesse U10), which is used for frequency locking of the lasers operating at 397, 422, 854, and \SI{866}{\nm}. The \SI{729}{\nm} $\mathrm{Ca}^+$ logic laser (not depicted in Fig.~\ref{fig:ca-lasers}) is locked to an ultra-stable reference laser at \SI{1542}{\nm} via a frequency comb \cite{scharnhorst_high-bandwidth_2015} and sent through a double-pass (DP) AOM for frequency tuning by $\pm50\,\mathrm{MHz}$.

The entire optical setup has been built on rigid breadboards with honeycomb core, which are interconnected via optical fibers. These breadboards will be stacked and installed in a standardized cooled 20-foot shipping container together with the remaining setup to make the entire system transportable.

\section{Characterization results}
\label{sec:experiment}
The following experiments on a single $^{40}\mathrm{Ca}^+$ ion were conducted with $\approx2.2\,\mathrm{W}$ of rf drive power to the helical resonator, which yields radial trapping frequencies of $\omega_\mathrm{rad,1}=2\pi\times2.65\,\mathrm{MHz}$ and $\omega_\mathrm{rad,2}=2\pi\times2.42\,\mathrm{MHz}$. The axial confinement in the 1~mm long experiment zone results in a trapping frequency of $\omega_\mathrm{ax}=2\pi\times1.64\,\mathrm{MHz}$.

\subsection{Imaging system}
\label{subsec:imsys}
The imaging system was designed for high-fidelity state discrimination with detection times well below a millisecond. This requires capturing a large solid angle fraction of emitted photons from a trapped ion and near-diffraction limited performance. The second design goal was the detection of individual ions in a linear string to allow the implemention of a multi-ion quantum logic clock \cite{schulte_quantum_2016}, which requires a sufficient optical magnification for typical ion-ion distances of a few micrometers and a field of view of around \SI{100}{\um}. 

We chose a single biaspheric lens due to its simplicity in assembly and alignment over more traditional multi-element lenses \cite{alt_objective_2002}. Viewports protruding into the vacuum enabled installation of the lens outside the vacuum chamber without compromising on the minimum distance $d_\mathrm{min}=36.6\,\mathrm{mm}$ of the first lens surface from the trap center. While a large lens diameter is advantageous in terms of the achievable numerical aperture and solid angle coverage, it can restrict optical access for diagonal laser beams (see Fig.~\ref{fig:achamber}). As a compromise, an outer lens diameter of \SI{50.8}{\mm} was chosen, leading to a 42.6~mm effective optical diameter and a numerical aperture $\mathrm{NA}=0.51$. A single monochromatic biaspheric lens can be optimized to minimize spherical and chromatic aberrations below the diffraction limit for a single point in the object plane. However, a trade-off between diffraction-limited performance and field-of-view needs to be made. We have therefore chosen to optimize the imaging performance for a circular field of view with \SI{100}{\um} radius.

The imaging system was numerically optimized using a commercial ray tracing software (Lambda Research OSLO), where the spheric coefficients, conic constants, and the curvature of the ASL surface facing the camera were free parameters. The damped least square optimization was carried out for an error function resembling the root mean squared image spotsize for three point sources separated by a maximum of 0.2~mm in the object plane. The resulting parameters are listed in Tab.~\ref{Tab_lens}. The simulation indicated that the correct adjustment of the lens-to-ion distance is critical for diffraction-limited performance.

\begin{figure}[htb!]
\includegraphics[width=8.0cm]{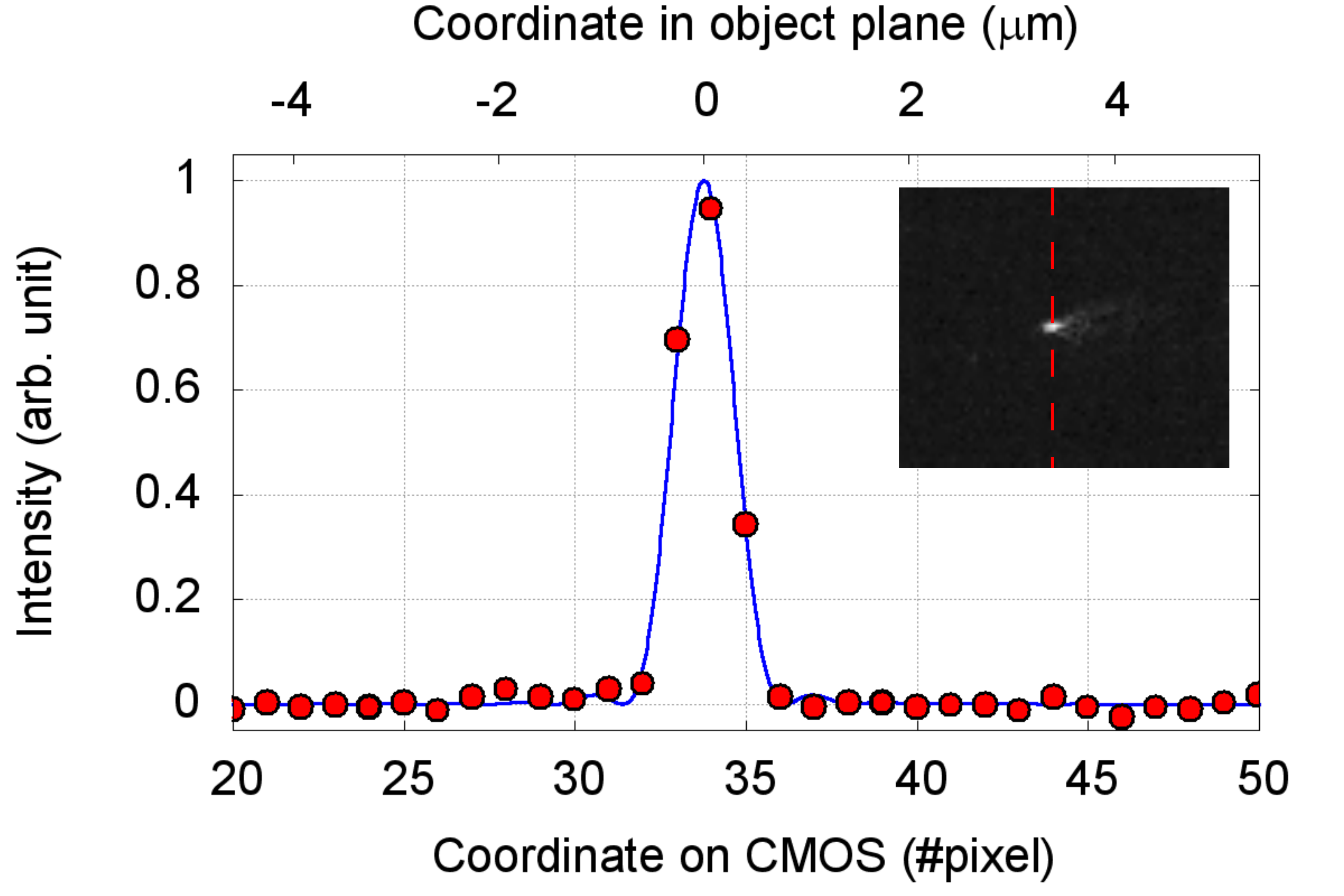}
\caption{Point spread function of a single $\mathrm{Ca}^+$ on the sCMOS chip (lower x-axis). Insert: photo of a single \ca ion. Red: measured intensity profile along the dashed line in the inset with electronic background substracted, blue: fit. The upper scale gives the corresponding dimensions in the object plane taking into account the magnification of the imaging system. Fitting the function $f(x)=I\cdot\left(\frac{2\cdot J_1\left(a\left(x-x_0\right)\right)}{a*\left(x-x_0\right)}\right)^2$ to the data resulted in $a=(1.62\pm0.03)\,\mathrm{pixel}^{-1}$ pixel, which is equivalent to \SI{0.793\pm0.013}{\um} in the object plane.}
\label{fig:psf-cam}
\end{figure}

Using a Doppler-cooled $\mathrm{Ca}^+$ ion loaded into the trap as the source, the ASL-to-ion distance and the CMOS to ASL distance were iteratively varied in order to minimize the ion`s point spread function (PSF) on the camera. Fig.~\ref{fig:psf-cam} shows an Airy function fitted to the count distribution on a straight line through the center of the PSF, depicted as red dashed line in the inset.
The first minimum of the fitted Airy function is found at \SI{0.793\pm0.013}{\um} from the global maximum. The value is given in the object plane, taking the 18-fold magnification of the imaging setup into account, which has been calibrated through an ion-ion distance measurement for known axial trapping frequencies.

This result is less than twice as large as the minimum resolvable distance of two point sources according to the Rayleigh criterion $d=0.61\lambda/NA\approx 475\,\mathrm{nm}$, where $\lambda=\SI{397}{\nm}$ is the wavelength and $NA=0.51$ the numerical aperture of the imaging system.

\begin{figure}[htb!]
\includegraphics[width=8.0cm]{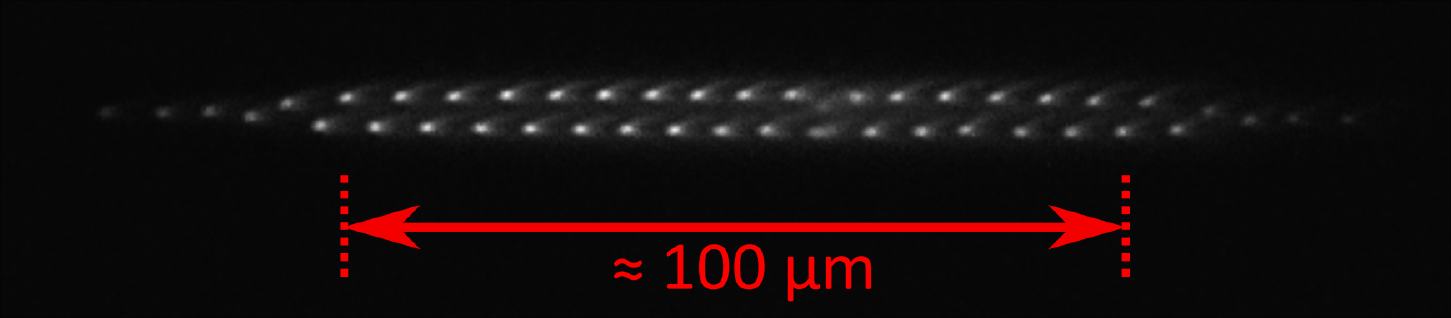}
\caption{Image of a 2d \ca ion crystal. The imaging performance is practically constant over a length scale of \SI{100}{\um}.}
\label{fig:large_crystal}
\end{figure}

We calculate the FWHM depth of field\cite{smith2000modern} to be $DOF=\pm\frac{\lambda}{2\cdot NA^2}=\pm \SI{0.765}{\um}$. Experimentally we determine $DOF<\SI{5\pm5}{\um}$, limited by the resolution of the employed precision translation stage. The biaspheric lens has been design for diffraction limited imaging in a field of view of \SI{100}{\um}. Fig.~\ref{fig:large_crystal} shows an image of a \ca ion crystal. A comparison of the PSFs of the central with the outer ions demonstrates that the imaging performance is practically constant over that length. The decrease in brightness towards the ends of the crystal is due to the limited diameter of the cooling laser beam.

We estimate the Strehl ratio\cite{Mahajan_83} (SR) for the biaspheric lens alone as
\begin{align}
SR=\exp\left(-\left(\frac{2\pi\sigma}{\lambda}\right)^2\right)=0.87,
\label{eq:strehl}
\end{align}
where we have taken $\sigma$ to be the total root-mean-squared form deviations of the aspheric surfaces measured by the manufacturer and $\lambda$ the wavelength. Wavefront aberrations of other optical elements in the imaging system, such as the vacuum windows and filters, will further reduce the Strehl ratio of the entire imaging system.

For a typical \SI{300}{\ms} exposure time of the sCMOS camera, we obtain $\approx11500$~counts in a $10\times10\,\mathrm{pixel}^2$ region of interest including the image of a \ca ion in the center with negligible contributions of scattered light from the cooling laser. The camera has a constant electronic offset of 99.3 counts with a rms noise of 1.9 counts in the absence of an ion, both numbers are given per pixel. 
Neglecting the constant offset, this results in an effective signal of $\approx1570$~counts and a signal to noise ratio of $\approx80$.

With the PMT we measure a dark count rate of below $0.1\,\mathrm{kHz}$ and a count rate of $\left(5.1\pm2.3\right)\,\mathrm{kHz}$ due to scattered light at 
\SI{397}{\nm}. For a single \ca we obtain a bright count rate of $\left(351\pm19\right)\,\mathrm{kHz}$, which is 55\% of the expected value of $\approx638\,\mathrm{kHz}$ at saturation, derived from 22.4~MHz linewidth divided by a factor of four arising from the equal distribution of the population over the $^2\mathrm{S}_{1/2}$, $^2\mathrm{P}_{1/2}$, and $^2\mathrm{D}_{3/2}$ levels in saturation \cite{wubbena_controlling_2014}, $7\%$ solid angle covered by the biaspheric lens, and a total transmission of $86\%$ through the optical system including the notch filter, the 90\,\% beam sampler reflectivity, and a photon counting efficiency of 0.3 specified by the manufacturer of the PMT. The reduction in count rate is compatible with aberration losses from specified wavefront errors in the optical imaging system, such as the Strehl ratio given in Eq.~\eqref{eq:strehl}.

\begin{figure}[htb!]
\includegraphics[width=8.0cm]{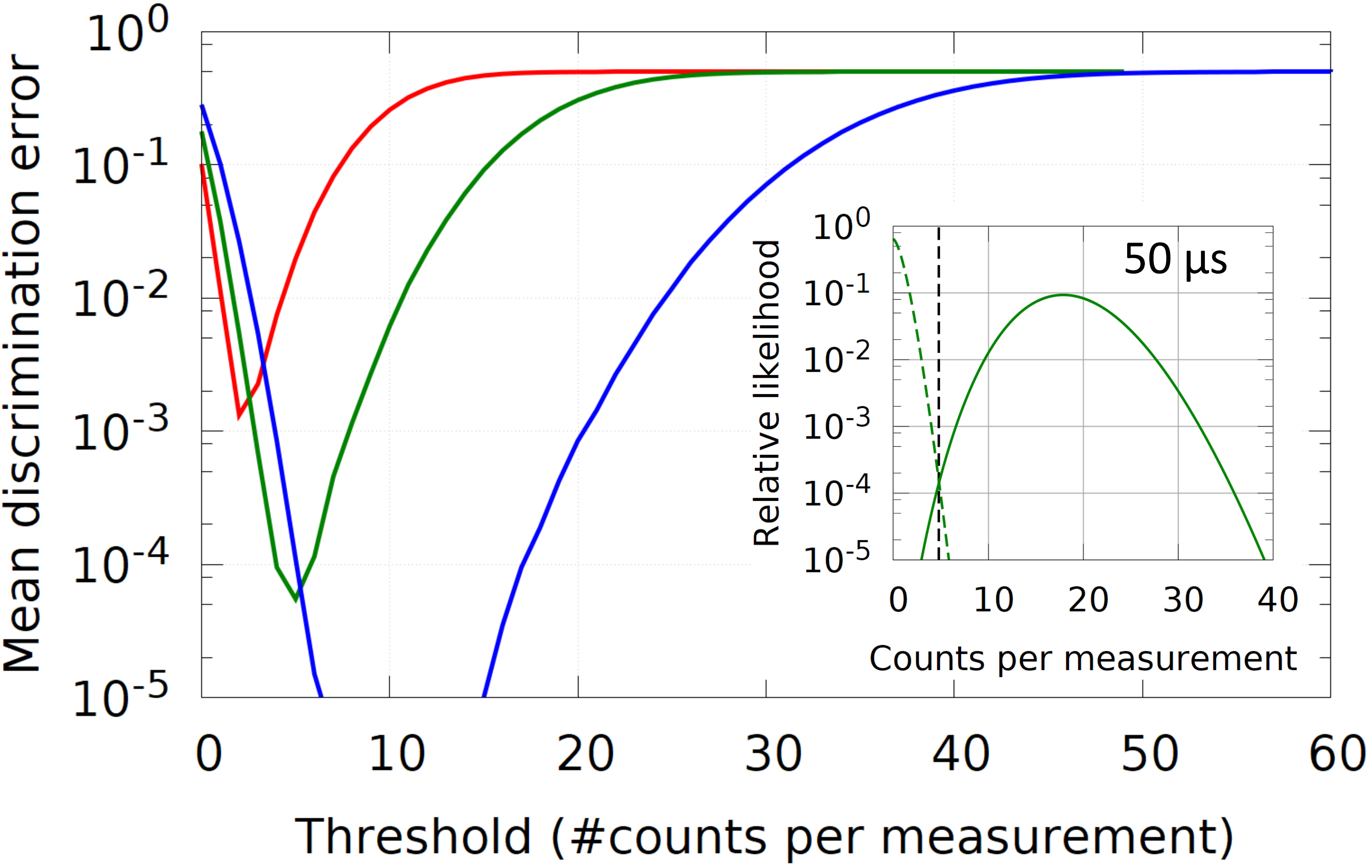}
\caption{State discrimination error for various detection times. The red (green/blue) curve shows the mean state discrimination error as a function of the discrimination threshold for \SI{25}{\us} (\SI{50}{\us} / \SI{100}{\us}) detection time. The inset shows the photon count distribution measured for \SI{50}{\us} detection time for a bright (dark) ion in red (blue) and a discrimination threshold of 5 counts per measurement indicated by the dashed black vertical line.}
\label{fig:derror}
\end{figure}

For state detection via electron shelving, it is required to distinguish between the ion being in the S$_{1/2}$ ground state (ion bright) and in the excited D$_{5/2}$ state (ion dark). We chose the threshold technique\cite{hemmerling_novel_2012} in which a threshold count is set above which the ion is declared bright. In Fig.~\ref{fig:derror} the mean state discrimination error is shown as a function of this threshold. The inset shows the measured photon count distributions for a bright (dark) ion in red (blue), which was prepared by applying (blocking) the \SI{866}{\nm} repumper laser during Doppler cooling. For the chosen detection time of \SI{50}{\us}, the two distributions partially overlap. Therefore any choice of the discrimination threshold (black dashed line) leads to a mean discrimination error of\cite{keselman_high-fidelity_2011}:
\begin{align}
\epsilon=\frac{p_\mathrm{b}(n\leq n_\mathrm{th})+p_\mathrm{d}(n> n_\mathrm{th})}{2},
\end{align}
where $n$ is the measured number of photons, $n_\mathrm{th}$ the threshold set, and $p_\mathrm{b}$ ($p_\mathrm{d}$) the probabilty to measure a bright (dark) ion. Thus, the numerator expresses the probability of a wrong state assignment to the measurement result.
In the main figure, this error is plotted as a function of the discrimination threshold for \SI{25}{\us} (\SI{50}{\us} / \SI{100}{\us}) detection time in red (green/blue). The red curve shows that a mean state discrimination error of below \SI{0.2}{\percent} for \SI{25}{\us} detection time can be achieved, which improves to below $10^{-5}$ for \SI{100}{\us} detection time.

\subsection{Wavelengthmeter frequency drift}
\label{subsec:wmlock}
The frequencies of the 397, 422, 854, and \SI{866}{\nm} are locked to a wavelength meter (WLM), which must therefore be continuously available.  The drift of the WLM limits how long it can run between recalibrations, which interrupt clock operation. While the general approach used here is similar to previous work \cite{couturier_laser_2018,saleh_frequency_2015,kobtsev_long-term_2007}, we employ single-mode fiber-switches connected via photonic crystal single-mode fibers to the WLM and do not perform any calibration of the WLM during the measurements.

To characterize the readout frequency drift, the frequencies of the 729 and \SI{397}{\nm} laser were recorded while both lasers were transfer-locked to a highly stable reference laser at \SI{1542}{\nm}\cite{scharnhorst_high-bandwidth_2015}, which typically drifts by less than 10~kHz per day. All devices were operated in a laboratory environment with $\pm0.2\,\mathrm{K}$ temperature stability during the measurement according to the temperature log of the air conditioning system. The WLM was placed on a vibration damped breadboard and covered by a non-insulating box to minimise air turbulence around the device. The temperature and air pressure measured by internal sensors of the WLM were recorded.

\begin{figure}[htb!]
\includegraphics[width=8.0cm]{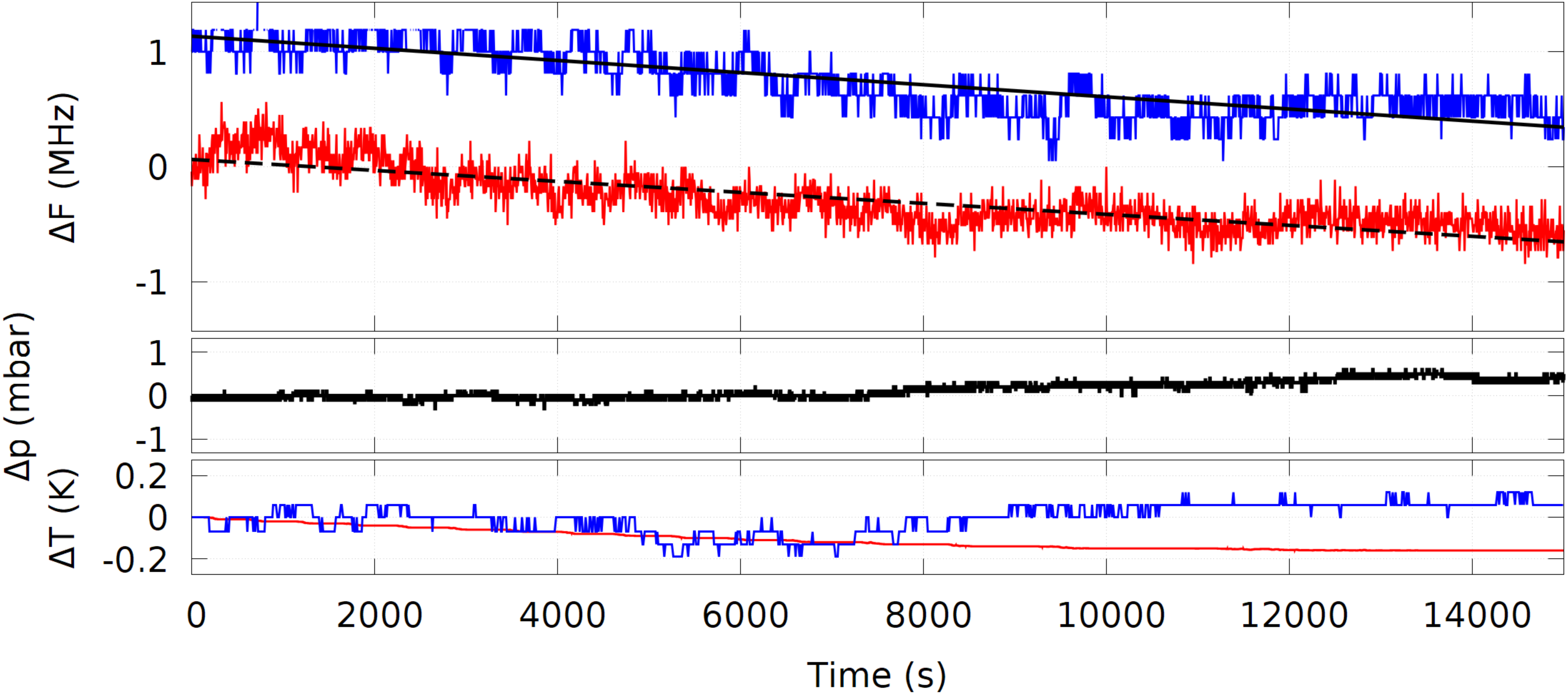}
\caption{Frequency drift of the 397 and \SI{729}{\nm} laser measured by the wavelength meter (WLM). Both lasers were transfer-locked to an ultrastable reference. Top: blue (red): measured \SI{397}{\nm} (\SI{729}{\nm}) frequency, black solid (dashed): linear fits. Middle: atmospheric pressure drift. Bottom: temperature measured by the WLM (red) and on top of the box covering the WLM (blue). All offsets have been removed. The linear fit results in a drift of $(-0.190\pm0.001)$~MHz/h ($(-0.171\pm0.001)$~MHz/h) for \SI{397}{\nm} (\SI{729}{\nm}). The residuals are normally distributed with a FWHM of approximately 0.4~MHz.}
\label{fig:wm-lock}
\end{figure}

The results are shown in Fig.~\ref{fig:wm-lock}. The measured frequency of the \SI{397}{\nm} (\SI{729}{\nm}) laser drifts by $-0.190$~MHz/h ($-0.171$~MHz/h). Both measurements fluctuate by approximately 1.0~MHz peak-to-peak on a timescale of one hour. The linear drifts are less than half of the values reported by Saleh \textit{et al.}\cite{saleh_frequency_2015} for a thermally isolated WLM from the same manufacturer but from another product series (WS-7). Moreover, the drifts reported here are more than two orders of magnitude smaller than the value obtained by Kobtsev \textit{et al.}\cite{kobtsev_long-term_2007} for another series (WS-8).
Since the drifts are small compared to the natural linewidth of the Doppler cooling and repumping transition, the WLM is well suited to lock the lasers listed above. The resonant Al ionisation laser could also be locked in this fashion, though we have not yet done so. This simplifies the setup compared to other solutions such as individual reference cavities for all lasers and is more cost efficient than a frequency comb\cite{scharnhorst_high-bandwidth_2015}.

\begin{figure}[htb!]
\includegraphics[width=8.0cm]{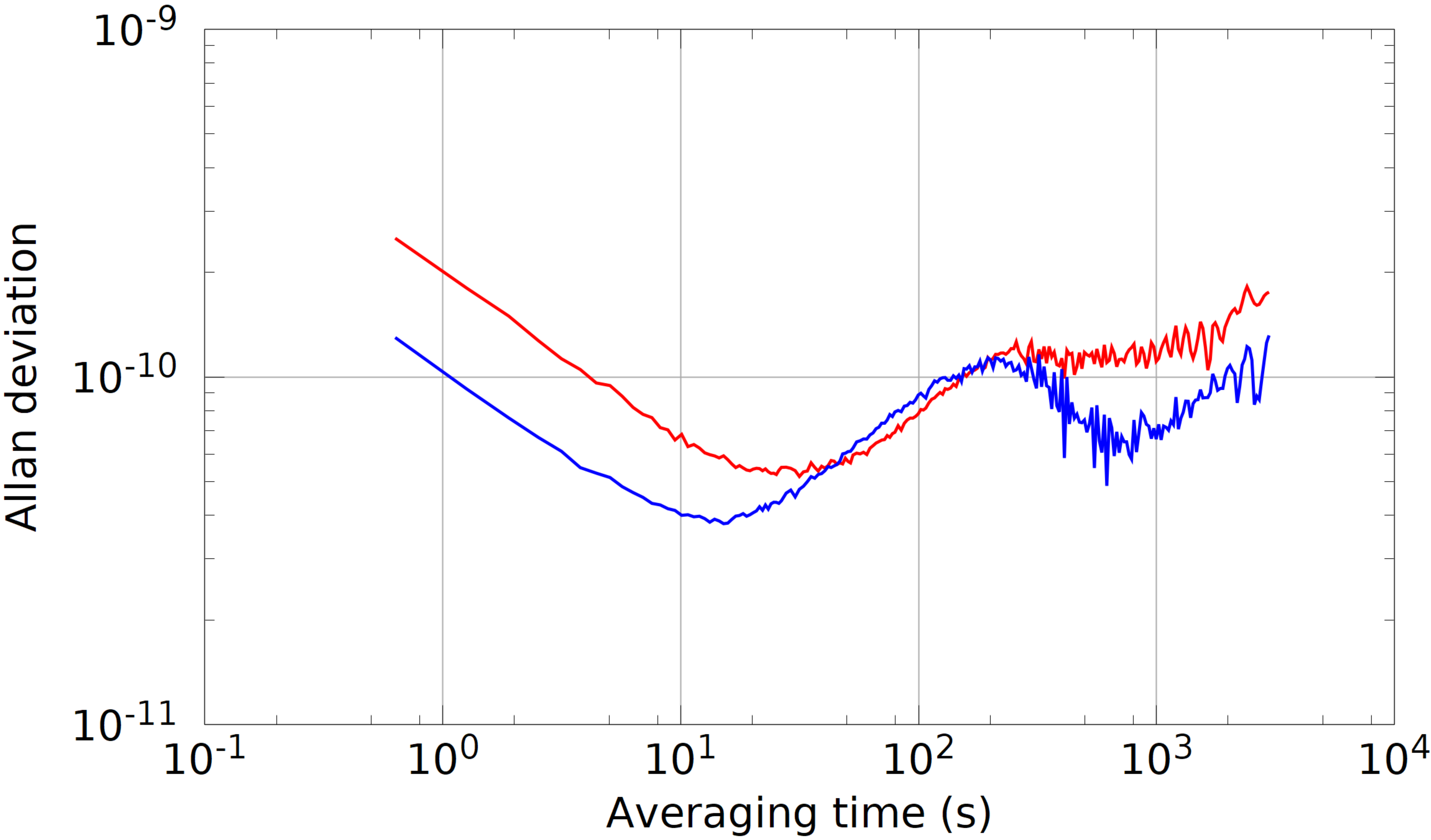}
\caption{The curves show the fractional frequency Allan deviation of the wavelength meter measurements at \SI{397}{nm} (blue) and \SI{729}{nm} (red). Both lasers were frequency stabilized to an ultrastable reference. The linear drift has been substracted. The measurements were taken simultanuously over \SI{15000}{\s}.}
\label{fig:wlm-adev}
\end{figure}

Fig.~\ref{fig:wlm-adev} shows the Allan deviations derived from the frequency recordings after substraction of the individual linear drifts. In both cases the Allan deviation stays below $3\times10^{-10}$ over the entire range of averaging times. This is comparable to the results measured by Saleh \textit{et al.}\cite{saleh_frequency_2015} and about a factor of three lower than the values reported by Courier \textit{et al.}\cite{couturier_laser_2018}.

\subsection{Ground state cooling}
\label{subsec:gsc}
The second-order Doppler shift on the \al clock transition is a significant contribution to the error budget, as will be discussed in Sec.~\ref{subsec:ebud}. It can be reduced by cooling the clock ion. Moreover, quantum logic spectroscopy requires ground state cooling of the motional mode used to transfer the electronic state of the clock ion to that of the logic ion\cite{schmidt_spectroscopy_2005}. The implementation of ground state cooling requires that the ion has a Lamb-Dicke factor $\eta<1$ and to be in the resolved-sideband regime. Requiring $\eta<1$ corresponds to the recoil energy from a single absorbed or emitted photon being less than the energy spacing between harmonic oscillator levels in the trap. This ensures that the heating rate from dissipation via photon scattering can be made small enough to reach the ground state\cite{wineland_experimental_1998}. In the regime $\eta\ll 1$, transitions that change the motional quantum number by more than 1 are strongly suppressed. In the resolved-sideband regime, the linewidth of the cooling transition is much smaller than the frequency of the motional mode to be cooled. In this situation, carrier transitions between electronic states that do not change the motional quantum number and red (blue) sideband (RSB/BSB) transitions that reduce (increase) the motional state by one quantum, can be spectrally addressed by a sufficiently narrow laser. Phonons can be removed from the harmonic motion by driving a RSB followed by a dissipative electronic state reinitialisation. We chose to implement pulsed sideband cooling\cite{roos_quantum_1999} on the S$_{1/2}$-D$_{5/2}$ transition with repumping via the P$_{3/2}$ state.
For typical radial (axial) trapping frequencies of $\oradca\approx2\pi\times2.5\,\mathrm{MHz}$ ($\oaxca\approx2\pi\times1.5\,\mathrm{MHz}$) and an angle of $60^\circ$ ($45^\circ$) between the mode to be cooled and the 729~nm beam, we obtain $\eta_\mathrm{rad}\approx0.04$ and $\eta_\mathrm{ax}\approx0.06$.

\begin{figure}[htbp!]
\includegraphics[width=8.0cm]{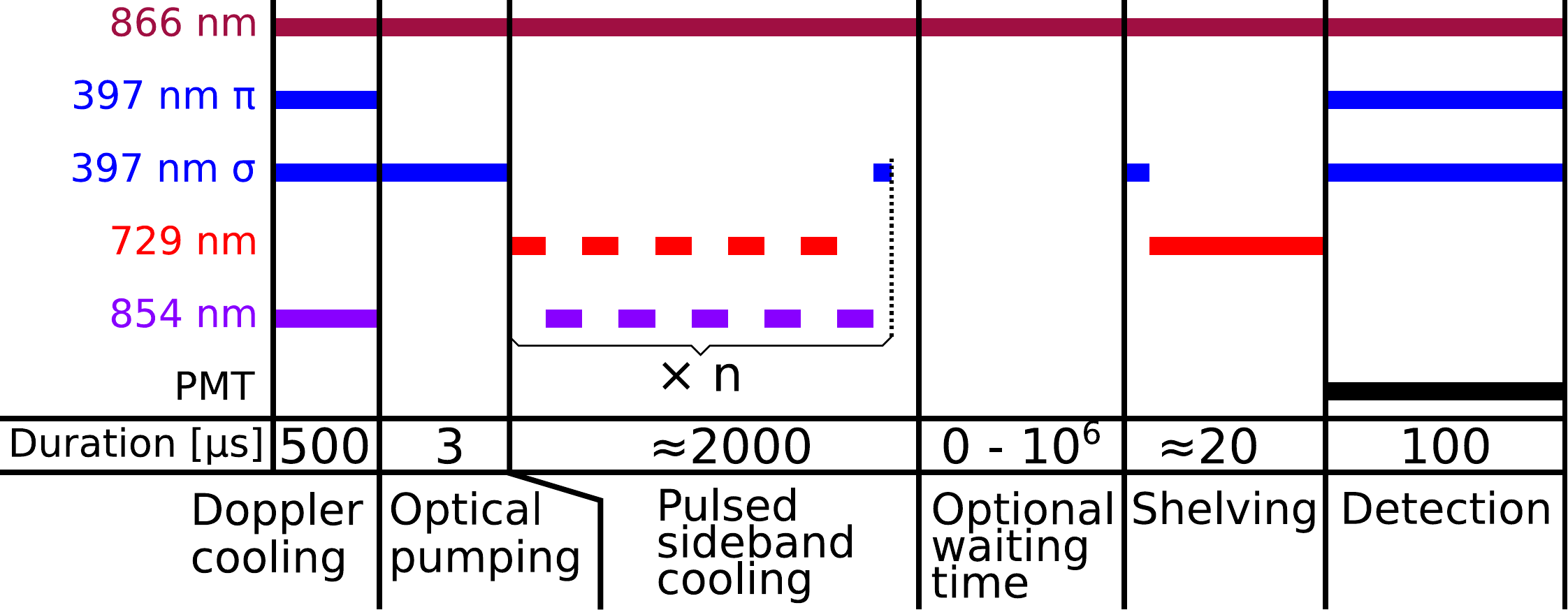}
\caption{Sideband cooling sequence. Colored lines represent active lasers. The optional waiting time is required for heating rate measurements.}
\label{fig:sbc_seq}
\end{figure}

Fig.~\ref{fig:sbc_seq} shows the experimental sequence starting with a \SI{500}{\us} long Doppler cooling pulse 10~MHz red detuned from the carrier, followed by \SI{3}{\us} optical pumping to the $^2\mathrm{S}_{1/2}, m_J = -1/2$ ground state. Then a series of interleaved \SI{729}{\nm} red sideband and \SI{854}{\nm} clearout pulses cools the ion near the motional ground state. The \SI{729}{\nm} pulses are approximate RSB $\pi$-pulses for $n=1$ to $n=0$ on the $\Delta m=-2$ transition, since repumping the D$_{5/2}, m_J=-5/2$ state via the P$_{3/2}, m_J=-3/2$ state only populates the initial state. The latter choice ensures a closed cooling cycle. A \SI{3}{\us} optical pumping pulse after every fifth clearout pulse restores the preparation to the electronic ground state $^2\mathrm{S}_{1/2}, \mathrm{m}_\mathrm{S} = -1/2$. An optional waiting time for heating rate measurements, electron shelving on the \SI{729}{\nm} transition, and a final state detection phase conclude the sequence.

The mean occupation number $\overline{n}$ is determined from the excitation $I_\mathrm{RSB}$ $\left(I_\mathrm{BSB}\right)$ of the first order red (blue) sideband \cite{monroe_resolved-sideband_1995, turchette_heating_2000}, both probed with the $\pi$-time of the blue sideband to maximise SNR:
\begin{align}
\overline{n}=\frac{1}{\frac{I_{\mathrm{BSB}}}{I_{\mathrm{RSB}}}-1}
\end{align}
We use $\approx70$ sideband cooling pulses, the minimum number needed to reach the steady state occupation $\overline{n}\approx0.05$.

\begin{figure}[htbp!]
\includegraphics[width=8.0cm]{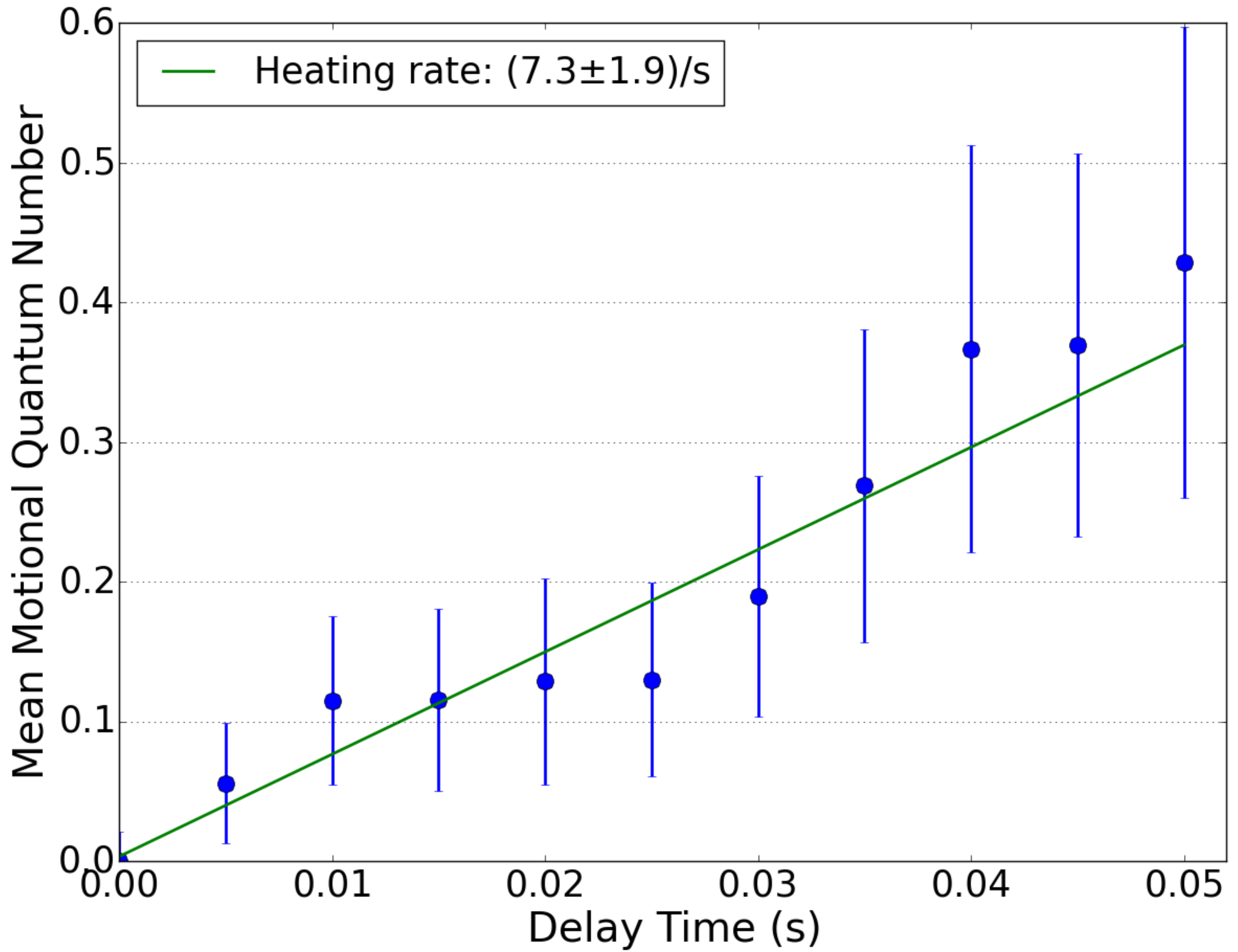}
\caption{Example of a heating rate measurement in the axial direction. Shown is the average motional quantum number as a function of the waiting time in the sideband cooling sequence (see Fig.~ \ref{fig:sbc_seq}) and a linear fit to the data (green). The error bars are derived from the statistical uncertainty of the sideband excitation measurements.}
\label{fig:sbc_heatrat}
\end{figure}

For the heating rate measurements, the waiting time $t$ in the sideband cooling sequence was varied between 0 and 200~ms. A linear function fit to the obtained $\overline{n}(t)$ yields the heating rate, as shown in Fig.~\ref{fig:sbc_heatrat} for the axial mode. Tab. \ref{Tab_gsc} summarizes the heating rates and temperatures for all three motional modes. The final $\bar{n}$ is limited by unsuppressed high frequency noise on the \SI{729}{\nm} laser, which leads to off-resonant carrier excitation during the red sideband pulse\cite{scharnhorst_high-bandwidth_2015}.

\begin{table} [htbp!]
\caption{Single $\mathrm{Ca}^+$ ground state motional quantum numbers, heating rates after sideband cooling and micromotion compensation, and electric field noise power spectral density at the secular mode frequencies.}
\label{Tab_gsc}
\begin{small}
\begin{center}
\begin{ruledtabular}
\begin{tabular}{lllll} 
Direction & $\frac{\omega (\mathrm{MHz})}{2\pi}$ & $\overline{n}$ & $\dot{n}\,\left(\mathrm{s}^{-1}\right)$ & $S_\mathrm{E}(\omega_j) \left(\frac{\mathrm{V}^2}{\mathrm{m}^{2}\mathrm{Hz}}\right)$
\tabularnewline
\toprule
 axial: & $1.64$ & $0.06\pm0.02$ & $7.3\pm1.9$ & $(8.2\pm2.1)\times10^{-14}$  
\tabularnewline
 radial I: & $2.42$ & $0.04\pm0.03$ & $30\pm4$ & $(5.0\pm0.7)\times10^{-13}$  
 \tabularnewline
 radial II: & $2.65$ & $0.13\pm0.08$ & $30\pm8$ & $(5.5\pm1.5)\times10^{-13}$  
 \tabularnewline
\end{tabular}
\end{ruledtabular}
\end{center}
\end{small}
\end{table}

\subsection{Micromotion compensation}
\label{subsec:mimoc}
Micromotion of ions induced by the confining rf-field causes second-order Doppler shifts whose uncertainties can significantly contribute to the error budget of an optical \al clock \cite{chou_frequency_2010}. While intrinsic micromotion is an unavoidable feature of rf traps, so-called excess micromotion (EMM) arising from phase differences in the applied rf field, stray electric fields that push the ion away from the nodal line of the radial electric quadrupole field, and imperfections of the electrode geometry, can be minimised\cite{berkeland_minimization_1998, keller_precise_2015}.  
To obtain less than $1\times10^{-18}$ fractional second-order Doppler shift due to EMM for an $^{27}\mathrm{Al}^+$ ion clock, the residual micromotion fields have to be kept below $66\,\mathrm{V/m}$ over an axial distance equal to the spatial extent of an \alca crystal under typical operating conditions. Therefore, the trap wafers of the multi-layer trap have to be aligned on the \SI{10}{\um} and 0.1~mrad level\cite{herschbach_linear_2012}. The conductors for the trapping rf potentials have to be length- and capacity-matched to avoid a phase shift between the electrodes.
Following \cite{berkeland_minimization_1998}, we estimate a required length-matching of $0.1\,\mathrm{mm}$ to keep the  additional fractional second-order Doppler shift below $1\times10^{-18}$. Micromotion from stray dc fields can be strongly suppressed by applying dc voltages to trap electrodes that compensate the external field.

\begin{figure}[htbp!]
\includegraphics[width=8.5cm]{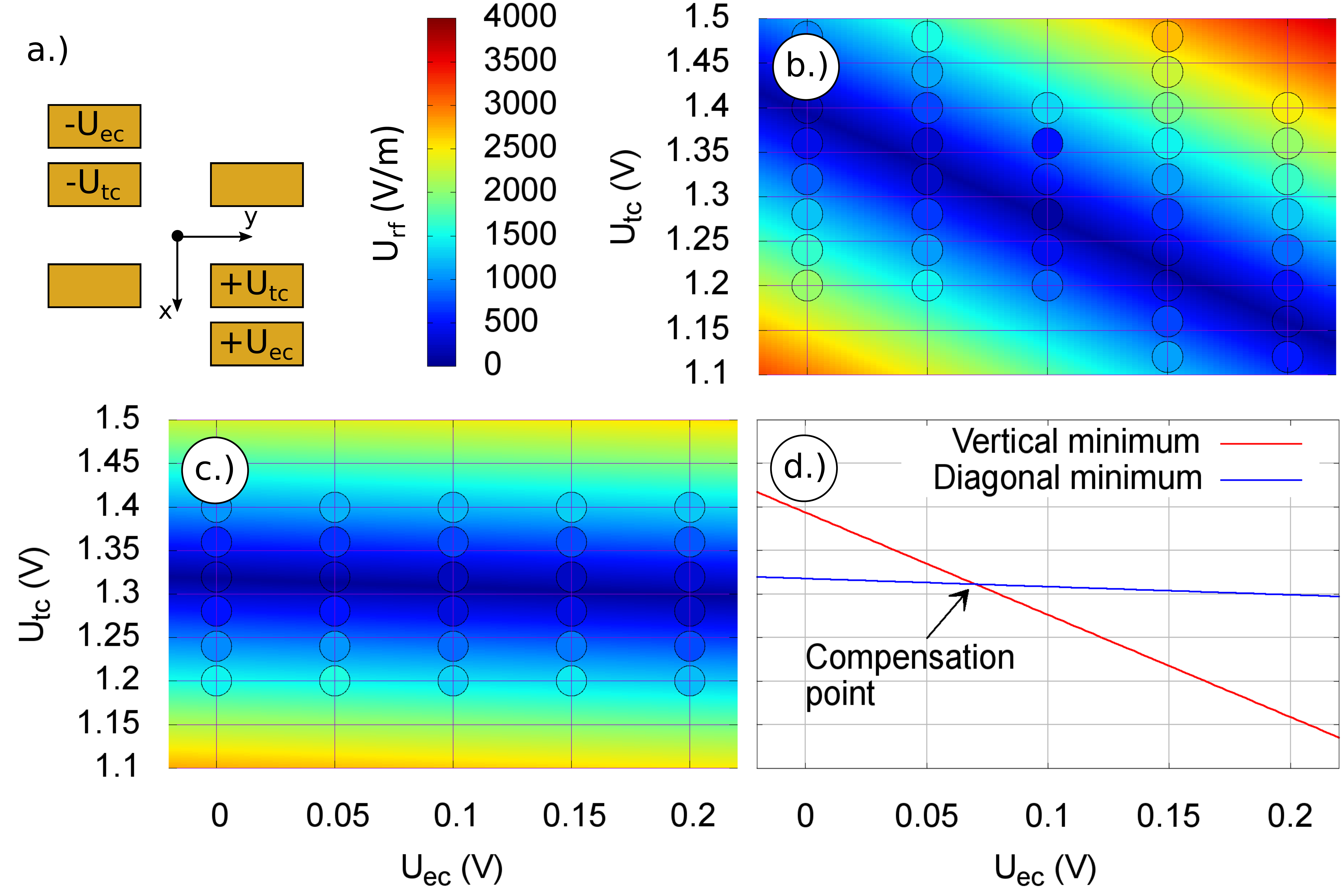}
\caption{Excess micromotion compensation in the plane defined by the vertical and diagonal \SI{729}{\nm} beam. a.) Schematic cross section of the trap showing the assignment of the voltages to the electrodes in the experiment zone. b.)/c.) vertical/diagonal rf field in \SI{}{\V\per\m} at the ion position as a function of the two compensation voltages. Filled dots represent the measured fields, the background depicts a v-shaped profile fit to the data. d.) Determination of the micromotion compensation point at the intersection of the two lines of minimum micromotion.}
\label{fig:mimoc-rad1}
\end{figure}

\begin{figure}[htbp!]
\includegraphics[width=8.5cm]{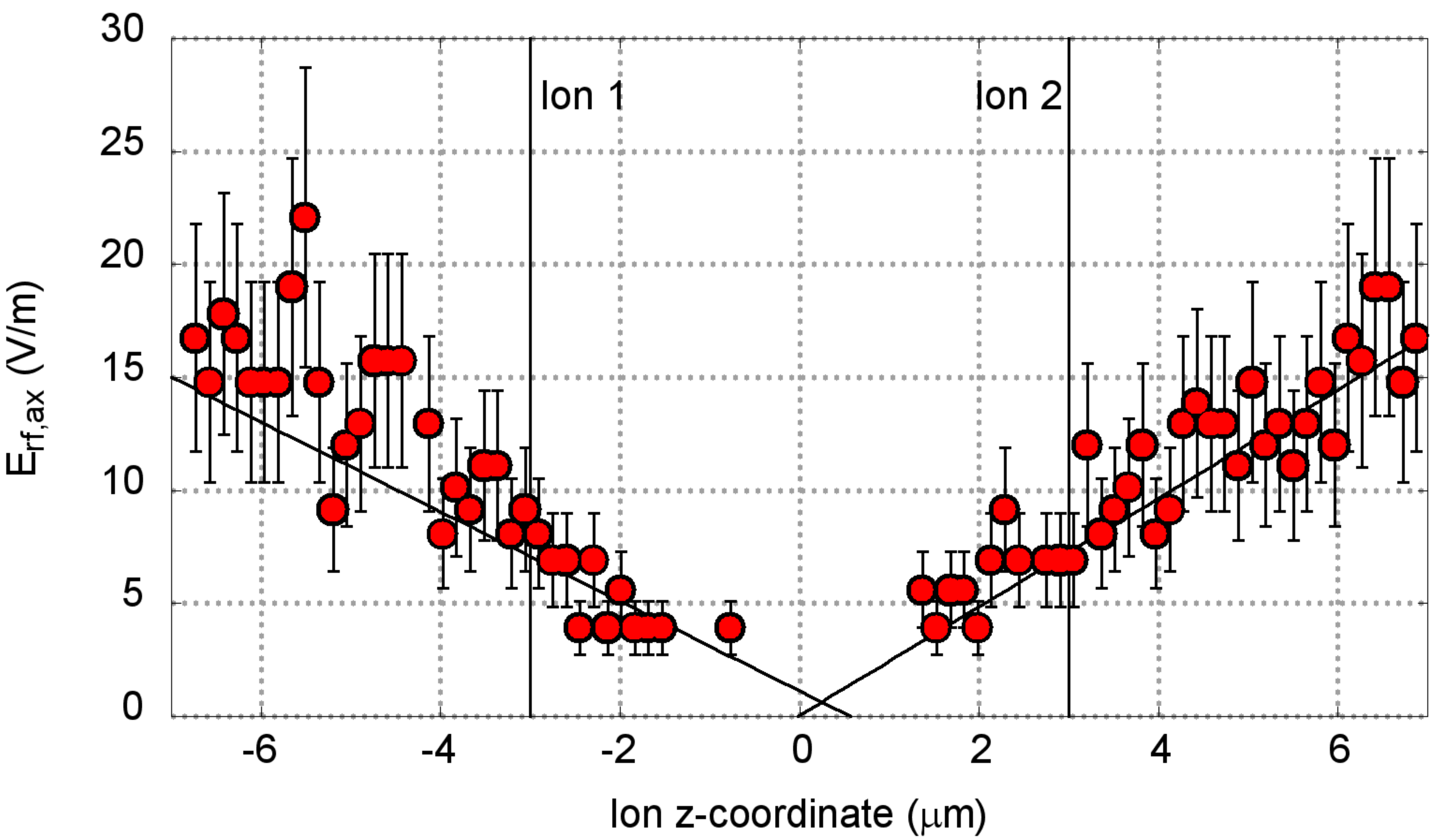}
\caption{Rf field along the axial direction. Shown is the axial component of the amplitude of the trap drive rf electric field as a function of the ion position along the trap axis. Two vertical lines indicate the possible positions of an \al ion in a two ion crystal placed symmetrically around the trap center for a relatively weak axial confinement of $\oaxca\approx2\pi\times\SI{0.9}{\MHz}$.}
\label{fig:mimoc-ax}
\end{figure}

Intrinsic and residual EMM can be measured using several techniques \cite{berkeland_minimization_1998, keller_precise_2015}. Here, we employ the resolved-sideband method by comparing the excitation rate when driving an rf motional sideband of the ion's transition spectrum $(r_\mathrm{mmsb})$ to the corresponding rate $r_\mathrm{car}$ when driving the carrier, as quantified by the sideband modulation index, which is quantified by the sideband modulation index\cite{berkeland_minimization_1998}
\begin{align}
\label{eq:bi}
\beta_i\approx2\sqrt{\frac{r_\mathrm{mmsb}}{r_\mathrm{car}}}.
\end{align}
It is minimized for three non-coplanar \SI{729}{\nm} probe directions $i=xz, y, z$ (see Fig.~\ref{fig:achamber}) to ensure compensation in all directions. In a first step, the ion was moved axially by applying differential voltages to neighboring trap zones to minimize excitation on the axial micromotion sideband. Subsequently, the voltages on the compensation electrodes in the \SI{1.0}{\nm} long experiment zone were scanned to displace the ion radially while measuring the modulation index for the vertical ($y$) and diagonal ($xz$) \SI{729}{\nm} beam. For each of the three \SI{729}{\nm} beams, the rf field in the corresponding direction, as experienced by the ion, is related to  $\beta$, the ion mass $m$, the laser wavenumber $k$ and the charge $Q$ \cite{berkeland_minimization_1998}:
\begin{align}
\label{eq:erf}
E_{\mathrm{rf},i}=\beta_i\frac{m\cdot \Omega_{rf}^2}{Q\cdot k}.
\end{align}
Fig.~\ref{fig:mimoc-rad1}, a.) shows the assignment of the compensation voltages to the electrodes in the experiment zone. The measured fields are represented by filled dots in subfigure b.) and c.). The background color depicts the result of fitting a v-shaped profile to the data. In subfigure d.) the set of compensation voltages for lowest micromotion in the plane spanned by the vertical and diagonal \SI{729}{\nm} laser is determined to be $U_\mathrm{ec}=(0.070\pm0.003)\,\mathrm{V}$, $U_\mathrm{tc}=(1.311\pm0.002)\,\mathrm{V}$ by intersecting the two resulting lines of minimum micromotion for the two directions. In the last step the axial compensation was repeated while the radial compensation voltages were kept constant, which yielded the field distribution shown in Fig.~\ref{fig:mimoc-ax}.

The residual rf fields at the trap center after EMM compensation are listed in Tab. \ref{Tab_params}. The field in the $x$-direction was inferred from the results in the $z$ and $xz$ directions by averaging over all possible phase shifts between those two components. The large uncertainty in the shift is a consequence of the unkown phase relation between the fields, which could be overcome by using e.g. the photon correlation technique \cite{keller_precise_2015}. Possible reasons for the non-vanishing rf field are a phase difference between the two rf electrodes due to a difference in length or capacitance of the rf conductors, or trap asymmetries arising from the alignment inaccuracy in assembly. 

The findings are comparable with the results obtained with a $^{172}\mathrm{Yb}^{+}$ in a trap of the same type\cite{pyka_high-precision_2014}, measured using the same technique.

For typical trap parameters, the two ions of an \alca crystal are spaced \SI{4.3}{\um} apart from each other symmetrically around the minimum of the axial dc potential. When the dc potential minimum and the rf potential minimum do not coincide, the \al ion is exposed to a stronger rf field than in the rf minimum. However, Fig.~\ref{fig:mimoc-ax} shows that the difference is smaller than \SI{10}{\V\per\m} and therefore does not substantially change the total rf field of ${17.4}^{+{8.2}}_{-{4.2}}\SI{}{\V\per\m}$ at the ion position, cf. Tab.~\ref{Tab_params}.

\begin{table} [htbp!]
\caption[Residual rf field at the trap center.]{Residual rf field at the trap center measured with a single $\mathrm{Ca}^+$ ion after excess micromotion compensation.}

\label{Tab_params}
\begin{small}
\begin{center}
\begin{ruledtabular}
\begin{tabular}{ll} 
\textbf{Direction} & $E_\mathrm{rf} (\SI{}{\V\per\m})$
\tabularnewline
\toprule
 trap axis ($z$): & $7.1\pm1.6$
\tabularnewline
 vertical ($y$): & $7.9\pm1.7$
 \tabularnewline
 diagonal ($xz$): & $9.8\pm3.0$
 \tabularnewline
 optical axis ($x$, averaged over phase): & ${13.9}^{+{8.5}}_{-{7.0}}$
  \tabularnewline
 quadrature sum of $x$ (avg), $y$, and $z$: & ${17.4}^{+{8.2}}_{-{4.2}}$
 \tabularnewline
\end{tabular}
\end{ruledtabular}
\end{center}
\end{small}
\end{table}

\subsection{Systematic error estimation}
\label{subsec:ebud}
The properties of the ion trap and surrounding vacuum system as measured by a single \ca ion can be used to estimate an error budget for an \alca clock in the setup presented here, as shown in Tab. \ref{Tab_shift}. For this, mass dependent shifts are scaled with the mass ratio of $^{27}\mathrm{Al}^+$ and $^{40}\mathrm{Ca}^+$.

We will first consider the contribution from the second-order Doppler or time-dilation shift from secular motion. It is directly related to the total kinetic energy $E_{\mathrm{kin}}$ in all secular motional modes according to
\begin{align}
\frac{\delta\nu_\mathrm{sec}}{\nu}=-\frac{E_{\mathrm{kin}}}{mc^2}.
\end{align}

During clock interrogation with a probe time $T_p$, the motional heating of the ion in the trap increases the kinetic energy. From the measured heating rate of a single \ca ion, we can determine the power spectral density of electric field noise, $S_\mathrm{E}(\omega_j)$ at the trapping frequency $\omega_j$ \cite{turchette_heating_2000}:

\begin{align}
S_\mathrm{E}(\omega_j)=\frac{4m\hbar\omega_j}{Q^2}\dot{n}.
\end{align}


Assuming a homogeneous fluctuating electric field as the source of heating, we can calculate its impact on the modes of a 2-ion crystal, which is characterized by the amplitudes of each ion. From the measured average motional quantum numbers and heating rates we calculate\cite{scharnhorst_multi-mode_2018}
\begin{align}
\label{eq:2dsec}
\frac{\delta\nu_\mathrm{sec}}{\nu}=-\frac{\hbar}{2m_\mathrm{Al}c^2}\sum_j\omega_j\left(\overline{n}_j+\frac{1}{2}+\frac{T_p}{2}\cdot\dot{n}_j\right)\left(s^*_j+s_j\right)
\end{align}
with the sum over the modes $j$. We set $s^*_j=1$ for radial modes and $s^*_j=0$ for axial modes to account for the effect of intrinsic micromotion\cite{berkeland_minimization_1998}. The $s_j$ are scaling factors of order unity which correct for the different motional amplitudes of the two ions in a mixed-species crystal\cite{wubbena_sympathetic_2012}. Since the individual $\overline{n}_j$ and $\dot{n}_j$ were not measured for all six modes of the crystal yet, we assume $\dot{n}_j=30$ for all modes, which leads to an estimate of $\overline{n}_j+\frac{1}{2}+\frac{T}{2} \dot{n}_j\approx3.5\pm0.4$ when averaged over a \SI{200}{\ms} interrogation time without cooling, based on the results for a single \ca ion shown in Tab.~\ref{Tab_gsc}. For the same trap strength that resulted in  $\oradca\approx2\pi\times2.5\,\mathrm{MHz}$ and $\oaxca\approx2\pi\times1.5\,\mathrm{MHz}$, the six motional mode frequencies and $s_i$ listed in Tab.~\ref{Tab_alcamod} are obtained. Using Eq.~\eqref{eq:2dsec}, this yields $\frac{\delta\nu_\mathrm{sec}}{\nu}=(-12.0\pm1.2)\times10^{-18}$, given in Tab.~\ref{Tab_shift}.
However, for a full clock evaluation the temperature of each mode after interrogation needs to be carefully measured, taking into account possible non-thermal distributions \cite{chen_sympathetic_2017}.

\begin{table} [htbp!]
\caption{Mode frequencies and scaling factors for the \alca crystal.}
\label{Tab_alcamod}
\begin{small}
\begin{center}
\begin{ruledtabular}
\begin{tabular}{llll} 
Mode $j$ & $\omega_i$ & $s_j$ & $s^*_j$
\tabularnewline
\toprule
Axial in-phase & 1.62 & 0.97 & 0
\tabularnewline
Axial out-of-phase & 2.92 & 0.97 & 0
\tabularnewline
Radial I in-phase & 1.88 & 1.19 & 1
\tabularnewline
Radial I out-of-phase & 3.29 & 2.29 & 1
\tabularnewline
Radial II in-phase & 1.88 & 1.19 & 1
\tabularnewline
Radial II out-of-phase & 3.29 & 2.29 & 1 
\tabularnewline  
  
\end{tabular}
\end{ruledtabular}
\end{center}
\end{small}
\end{table}

Uncompensatable excess micromotion as discussed in Sec.~\ref{subsec:mimoc} results in a fractional second-order Doppler shift $\delta\nu_\mathrm{EMM}/\nu$. Following\cite{berkeland_minimization_1998} it is given by
\begin{align}
\label{eq:2demm}
\frac{\delta\nu_\mathrm{EMM}}{\nu}=-\frac{1}{2c^2}\sum_{i=x,y,z}\left\langle v_i^2\right\rangle,
\end{align}
where $c$ is the vacuum speed of light and $v_i$ are the components of the velocity vector derived from the trap drive E-field measurements:
\begin{align}
\label{eq:vi}
\left\langle v_i^2\right\rangle=\left(\frac{Q\cdot E_{\mathrm{rf,}i}}{\sqrt{2}\cdot\Omega_\mathrm{rf}\cdot m_\mathrm{Al}}\right)^2. 
\end{align}
By inserting Eq.~\eqref{eq:bi}, \eqref{eq:erf}, and \eqref{eq:vi} in Eq.~\eqref{eq:2demm}, we obtain \sods.

Another contribution to the error budget is the ac Stark shift due to blackbody radiation given by\cite{rosenband_blackbody_2006,dolezal_analysis_2015}:
\begin{align}
\frac{\delta\nu_\mathrm{BBR}}{\nu}&=-\frac{\pi\left(k_BT_\mathrm{env}\right)^4\Delta\alpha(0)}{60\epsilon_0\hbar^4c^3}\\
&=-3.8\times10^{-18}\cdot\left(\frac{T_\mathrm{env}}{300\,\mathrm{K}}\right)^4
\end{align}
where $\Delta\alpha(0)=(0.82\pm0.08)\times10^{-41}\mathrm{Jm}^2\mathrm{V}^{-2}$ is the calculated static differential polarizability of the \al clock transition\cite{safronova_precision_2011}, $\epsilon_0$ the vacuum permittivity, and $T_\mathrm{env}$ the temperature of the ion's thermal environment.

About \SI{40}{\percent} of the solid angle "seen" by the ion is covered by the vacuum chamber and the viewports, which are assumed to be at room temperature of approximately \SI{300}{\kelvin}. The temperature can be measured with an uncertainty of well below \SI{\pm1}{\kelvin} using standard thermistors placed on the outside of the chamber. The other \SI{60}{\percent} are covered by the trap, which is heated due to dielectric rf loss. The chosen trap drive frequency of \SI{24.65}{\MHz} and amplitude of \SI{885}{\V} yield a maximum temperature of about \SI{5}{\kelvin} above its environment\footnote{The trap chip temperature measurement was performed by the Czech metrology institute (CMI).}. As a conservative estimate, an averaged environmental temperature of $T_\mathrm{env}=\SI{302.5\pm2.5}{\kelvin}$ was assumed for the entire environment, i.e. the vacuum chamber and the ion trap. This results in $\delta\nu_\mathrm{BBR}/\nu=(-4.0\pm0.4)\times10^{-18}$, where the uncertainty is dominated by $\Delta\alpha$.

The last contribution treated here is the shift caused by collisions with background gas particles during clock interrogation. Only Langevin collisions lead to a phase shift large enough to be considered in current optical clocks. The frequency shift $\Delta\nu_c$ can be calculated by multiplying the collision rate $\Gamma_c$ with a scaling factor \cite{vutha_collisional_2017} of 0.16. Since this estimate assumes a worst case phase shift of $\frac{\pi}{2}$ for every collision event \cite{rosenband_frequency_2008}, it provides an upper limit to the uncertainty of the shift rather than the shift itself. The collision rate was measured by observing 21 reorderings of a \alca crystal during a total observation time of \SI{3892.5}{\s}. Assuming that a reordering takes place for every second collision event, a collision rate of $\Gamma_c\approx\SI{0.0108}{\per\second}$ and a fractional frequency uncertainty of $1.5\times10^{-18}$ are obtained. The uncertainty of the collision shift could be further reduced by detecting collision events through reduced fluorescence of the hot $\mathrm{Ca}^+$ ion \cite{wubbena_sympathetic_2012} and discarding the corresponding data points.

First order Doppler shifts due to motion of the trap relative to the clock laser\cite{rosenband_frequency_2008}, perhaps as a result of thermal expansion of the apparatus, could be mitigated in the future by setting up an interferometer using the mirrors on the trap wafers to phase-stabilize the clock laser to the position of the trap. 

\begin{table} [htbp!]
\caption{Estimated partial error budget for PTB's transportable \alca clock}
\label{Tab_shift}
\begin{small}
\begin{center}
\begin{ruledtabular}
\begin{tabular}{lll} 
Shift & Value $\left(10^{-18}\right)$ & Uncertainty $\left(10^{-18}\right)$
\tabularnewline
\toprule
Black body radiation: & -4.0 & 0.4
\tabularnewline
\hline
Excess micromotion\\second-order Doppler: & -0.4 & $^{+{0.4}}_{-{0.3}}$
\tabularnewline
\hline
Secular motion\\second-order Doppler: & -12.0 & 1.2
\tabularnewline
\hline
Background gas\\Langevin collisions: & 0.0 & 1.5

\tabularnewline
  
\end{tabular}
\end{ruledtabular}
\end{center}
\end{small}
\end{table}

\subsection{Ablation loading of Aluminium ions}
\label{subsec:all}
\begin{figure}[htbp!]
\includegraphics[width=8.0cm]{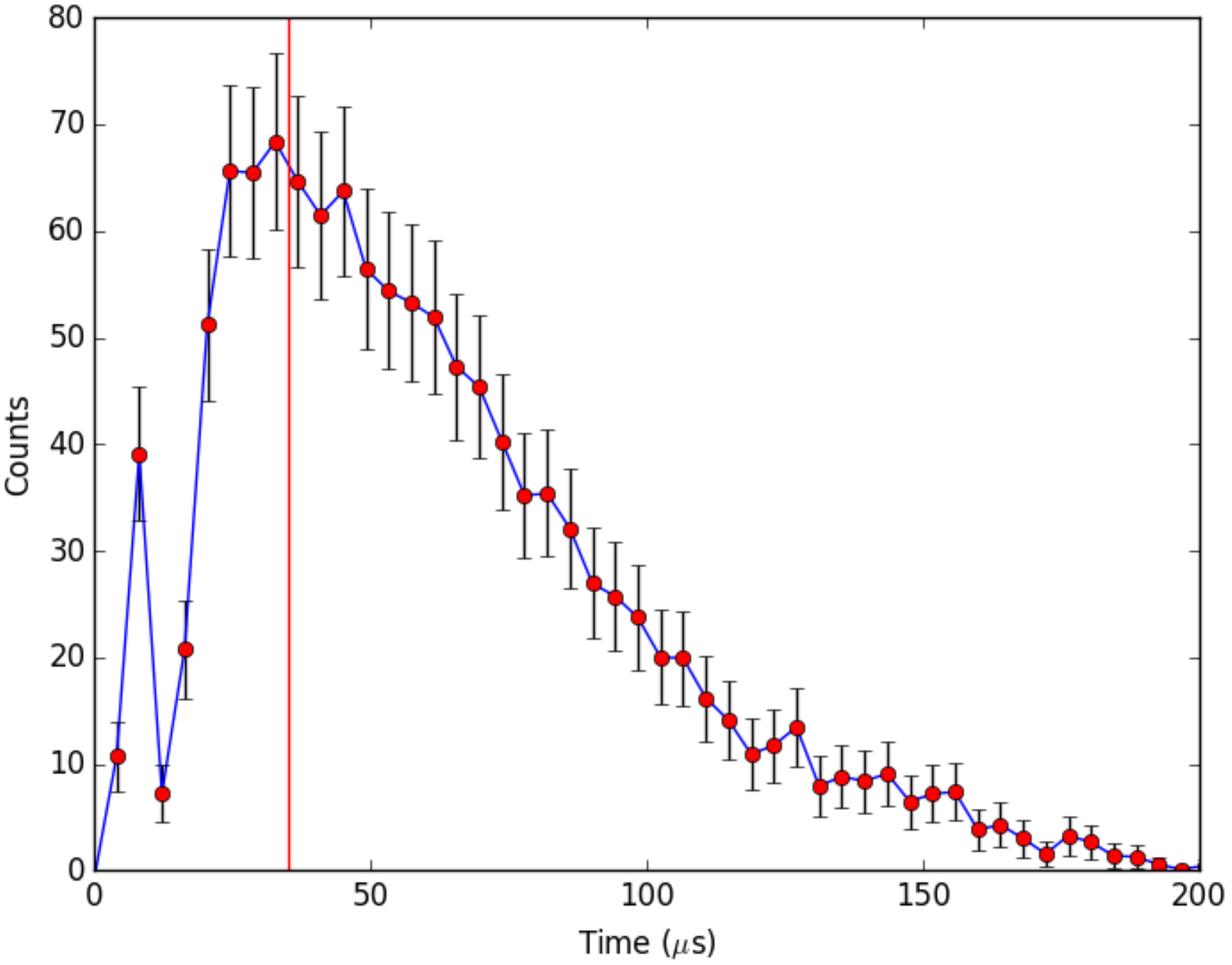}
\caption{Aluminium time of flight spectrum measured at the center of the loading zone. The counts are binned in \SI{2}{\us} time frames and the errorbars show the statistical error per bin. The red vertical line marks the start of the photoionization pulse at \SI{35}{\us} and the blue line links the data points as guide to the eye.}
\label{fig:al-tof}
\end{figure}

While ablation loading of $\mathrm{Ca}^+$ is relatively straightforward, thanks to the comparatively low neutral atom velocities and the availability of direct laser cooling right after ionization, ablated $\mathrm{Al}$ atoms travel much faster and have to be cooled sympathetically.  Average cooling times to crystallization of around 10~minutes have been reported\cite{guggemos_sympathetic_2015}. One way to reduce the crystallization time of a newly loaded \al ion is to lower its initial kinetic energy, e.g. by velocity-selective ionization of slower neutral atoms. Fig.~\ref{fig:al-tof} shows the velocity distribution of the neutral $\mathrm{Al}$ atoms, obtained from time-of-flight measurements\cite{guggemos_sympathetic_2015} of the Al fluorescence with a resonant ionization laser on the $^2$S$_{1/2}$-$^2$P$_{3/2}$ transition at \SI{394}{\nm}. The initial peak at \SI{8}{\us} resembles the velocity distribution measured by Guggemos \textit{et al.}\cite{guggemos_sympathetic_2015} with a peak at \SI{4500}{\m\per\s}. The majority of the $\mathrm{Al}$ atoms observed in our setup, however, come in a broad distribution with a maximum at \SI{35}{\us} TOF, corresopnding to a much lower velocity of \SI{490}{\m\per\s}.

\begin{figure}[htbp!]
\includegraphics[width=8.0cm]{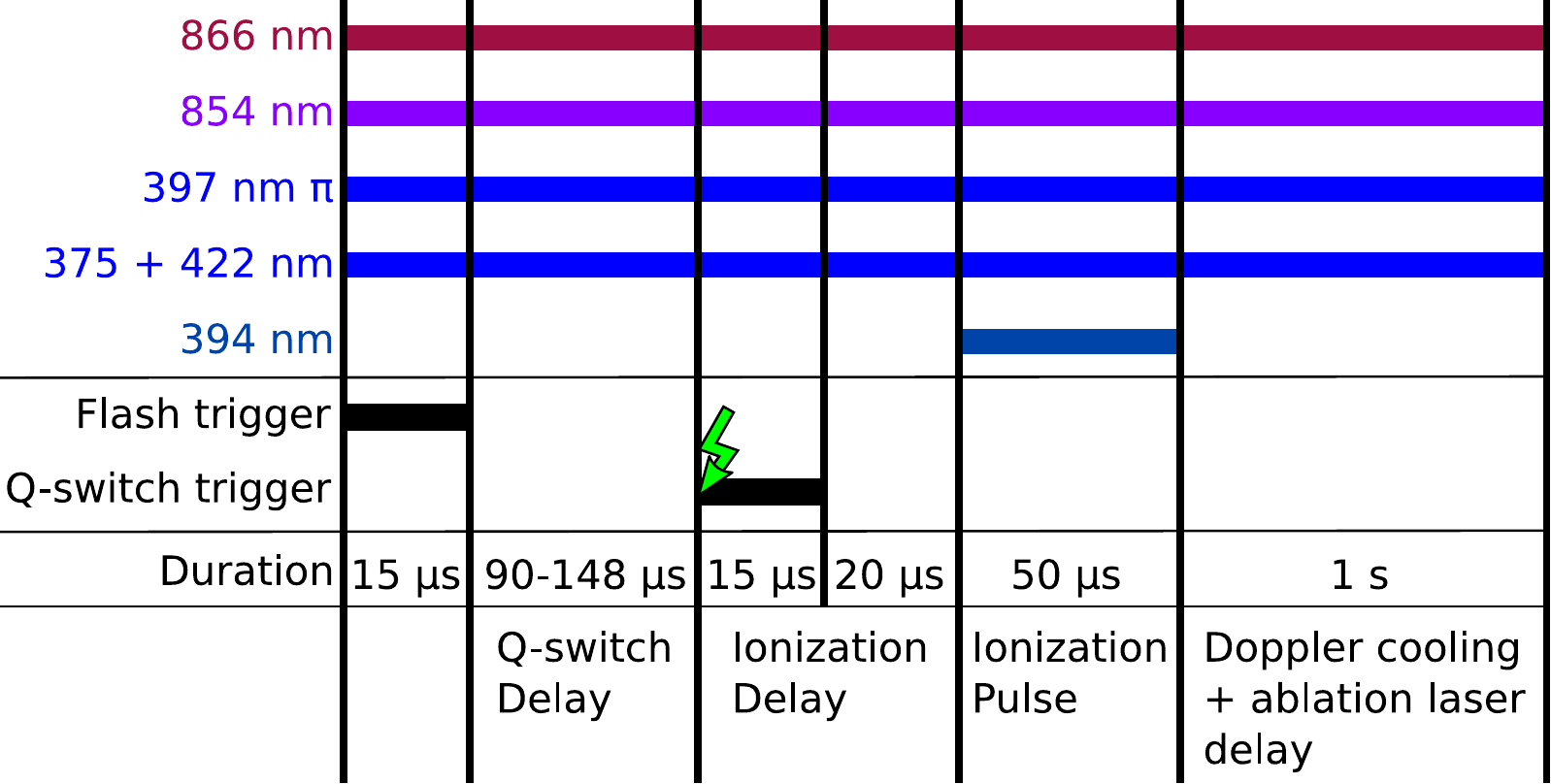}
\caption{$\mathrm{Al}^+$ loading sequence. The $\mathrm{Ca}^+$ ion is Doppler cooled during the entire sequence. First, the ablation laser flash lamp and its Q-switch are triggered by the experiment control system. The intermediate delay sets the pulse energy. At the rising edge of the Q-switch trigger pulse, the \SI{5}{\ns} long ablation laser pulse is fired (green flash symbol). After a second delay, the resonant Al ionization laser is unblocked to ionize relatively slow atoms, which have a longer travel time to the trap compared to fast atoms. To increase the probability of driving the non-resonant second ionization step, the Ca ionization lasers are switched on during the entire loading sequence. The sequence concludes with a wait time for the next available pulse of the ablation laser.}
\label{fig:al-load}
\end{figure}

For $\mathrm{Al}^+$ loading, the \SI{394}{\nm} beam is shuttered with a single pass AOM synchronized to the ablation laser pulse to select a slow velocity class of neutral atoms via time-of-flight selection in order to reduce the crystallization time. The loading sequence is depicted in Fig.~\ref{fig:al-load}, starting from a Doppler-cooled and crystallized $\mathrm{Ca}^+$ ion in the loading zone of the trap. First, the ablation laser flash lamp is triggered and after a delay, which can be adjusted to set the maximum instantaneous power of the pulse, a second trigger activates the Q-switch and therefore fires the ablation laser pulse (green flash symbol). After an additional delay starting with that pulse, the \SI{394}{\nm} laser is switched on for velocity selective $\mathrm{Al}$ ionization. A 1~s long Doppler cooling pulse follows to prevent exceeding the maximum repetition rate of the ablation laser in case multiple ablation shots are going to be fired. In order to increase the probability of successfully driving the non-resonant second step of the $\mathrm{Al}$ photoionization, both $\mathrm{Ca}^+$ ionization lasers are applied during the entire loading sequence.

\begin{figure}[htbp!]
\includegraphics[width=8.0cm]{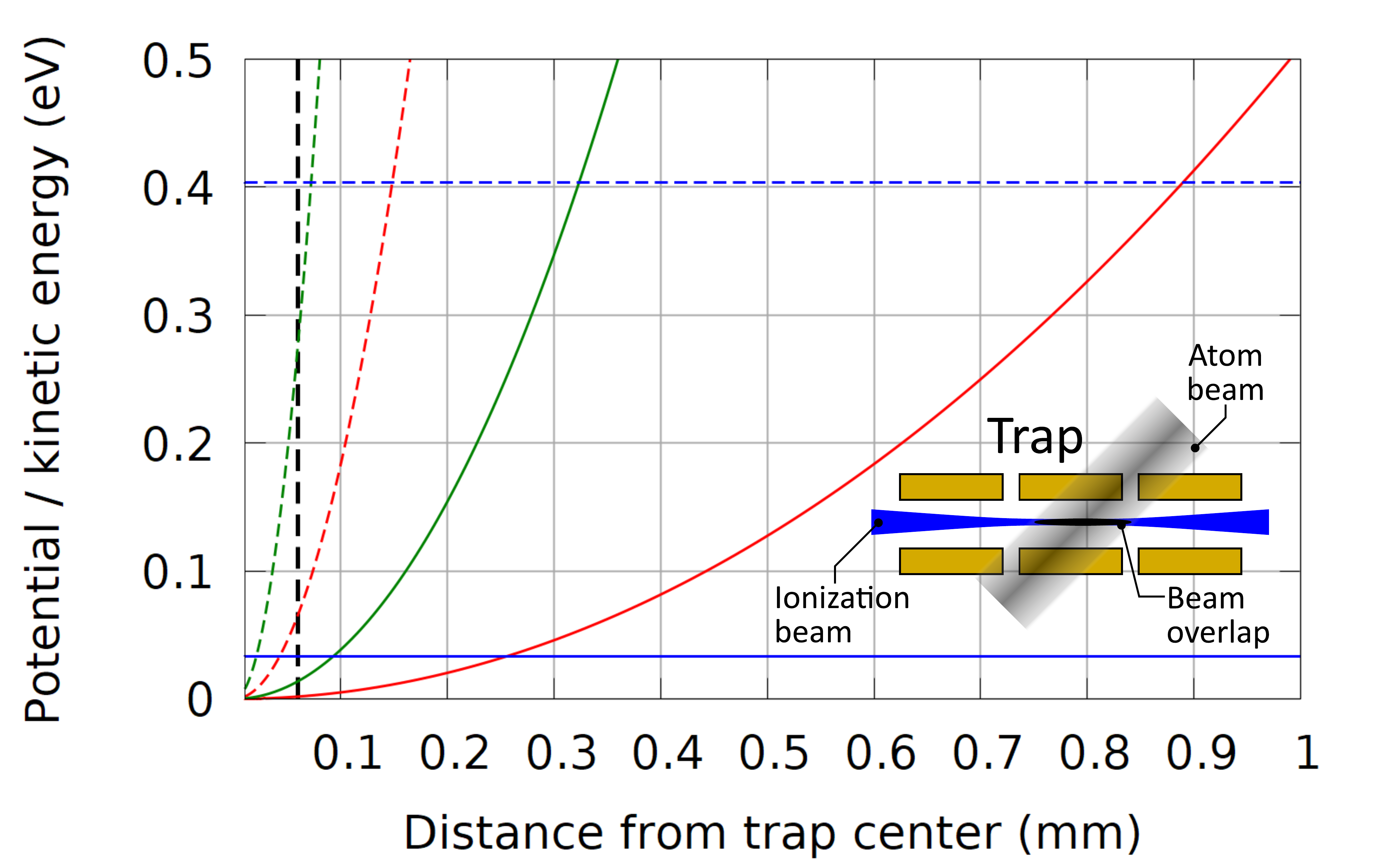}
\caption{Energy of \al as a function of the ionisation position in the trap (calculation). The horizontal solid (dashed) blue line indicates the kinetic energy of a neutral atom arriving \SI{35}{\us} (\SI{10}{\us}) after the ablation laser pulse. The solid (dashed) red curve shows the potential energy of \al as a function of displacement from the trap center for the standard trap potential (dashed) and for a relaxed potential used when loading (solid). The axial (green) trap frequencies are $\omega_\mathrm{z,Ca+}=2\pi\times0.25\,\mathrm{MHz}$ ($\omega_\mathrm{z,Ca+}=2\pi\times1.5\,\mathrm{MHz}$) for the two potentials while the radial trap frequencies are $\oradca\approx2\pi\times0.52\,\mathrm{MHz}$ ($\oradca\approx2\pi\times2\,\mathrm{MHz}$). The vertical dashed black line indicates the waist of the ionisation light beam from the LMA fiber, which restricts the volume where ionization takes place, cf. trap cross section in the inset.}
\label{fig:al-energies}
\end{figure}

The total energy of an \al ion depends on the position of ionization inside the trap as shown in Fig.~\ref{fig:al-energies}. The initial kinetic energy of a neutral atom arriving \SI{35}{\us} (\SI{10}{\us}) after the ablation laser pulse is shown by the solid (dashed) blue curve. Setting a photoionization delay of \SI{35}{\us} reduces the maximum kinetic energy of the neutral atoms by more than an order of magnitude to 0.033~eV, while keeping more than \SI{70}{\percent} of the atoms available for ionisation (cf. Fig.~\ref{fig:al-tof}). The volume in the loading zone, where the photoionization of a neutral Al atom takes place, is defined by the spatial overlap of the axial Al ionization lasers, as depicted in the inset. Its extent defines the maximum potential energy the atom gains by ionization away from the trapping potential minimum. The main figure shows the position dependence of the potential energy for two trapping configurations.  The dashed lines show the standard trap potential corresponding to $\omega_\mathrm{z,Ca+}=2\pi\times1.5\,\mathrm{MHz}$ in the axial (red) direction and $\oradca\approx2\pi\times2\,\mathrm{MHz}$ in the radial (green) direction.  The solid lines show a relaxed potential for which $\omega_\mathrm{z,Ca+}=2\pi\times0.25\,\mathrm{MHz}$ and $\oradca\approx2\pi\times2\,\mathrm{MHz}$. The potential energy gained due to a displacement from the trap center of less than \SI{50}{\um} in standard trapping configuration (dashed red and green curve) already exceeds the kinetic energy of the ablated neutral atom. But for the relaxed trap, the potential energy remains below the initial kinetic energy for the entire ionization laser beam cross section, as illustrated by the vertical dashed black line indicating the waist of this beam. Therefore, the total energy of \al right after ionization is dominated by the axial displacement from the trap center.

For the relaxed trap, we measure an average time of $(230\pm100)\,\mathrm{s}$ between the ablation laser pulse on the Al target and the formation of an \alca crystal, where the uncertainty is given under the assumption of normally distributed loading times. This is roughly a factor of 2.5 shorter than the value reported for the same combination of species by Guggemos \textit{et al.}\cite{guggemos_sympathetic_2015}, who worked at higher trap frequencies of $\oradca\approx2\pi\times2\,\mathrm{MHz}$ and $\omega_\mathrm{z,Ca+}\approx2\pi\times0.4\,\mathrm{MHz}$.

\section{Summary and Outlook}
\label{sec:sumout}
We presented PTB's transportable \alca optical clock setup. The system is robust, relatively compact, and simple, due to the full fiberization of the $^{40}\mathrm{Ca}^+$ repumper laser system, locking all lasers required for loading and cooling of $^{40}\mathrm{Ca}^+$ to a wavelength meter, and employing a single biaspheric lens for imaging. With its NA of 0.51, sub micrometer resolution is achieved by imaging a $^{40}\mathrm{Ca}^+$ ion on a compact CMOS camera. Since the imaging system covers a large solid angle fraction of \SI{7}{\percent}, \SI{10}{\percent} of the collected fluorescent light suffice to reach a signal-to-noise ratio of 80 for \SI{300}{\ms} exposure time of the CMOS chip. The remaining photons are directed onto a PMT for state discrimination. We measure a mean state discrimination error of below \SI{0.2}{\percent} for \SI{25}{\us} detection time, which drops to below $10^{-5}$ for \SI{100}{\us}.
The ions are confined in a segmented multi-layer trap, which is loaded via neutral atom ablation. Pulsed PI of Al using time-of-flight velocity selection enables crystallization of \al on average after 4~minutes. We compensated excess micromotion down to a residual rf electric field at the trap center of ${17.4}^{+{8.2}}_{-{4.2}}$~\SI{}{\V\per\m}. Ground state cooling of a single $\mathrm{Ca}^+$ has been achieved by pulsed sideband cooling and heating rates of less than 10~quanta/s have been measured for all three motional modes of a single $\mathrm{Ca}^+$ ion. Those allow for interrogation times of a few hundred milliseconds without the necessity for simultaneous sympathetic cooling in a future \alca quantum logic optical clock.
We estimate a preliminary apparatus-related partial error budget for an \alca clock operated in the setup. Taking into account the secular motion and excess micromotion second-order Doppler shift, the blackbody radiation shift, and the shift due to Langevin collisions with background gas particles during clock interrogation, we estimate a systematic fractional frequency uncertainty of $1.9\times10^{-18}$. This is equivalent to a chronometric leveling height resolution of below \SI{2}{\cm}. These values are comparable to the recently published results for laboratory neutral atom Yb lattice clocks \cite{mcgrew_atomic_2018}.

Once a \alca clock based on the system presented here is installed inside a climate-controlled  20-foot shipping container, it will allow for  chronometric leveling at geodetically relevant sites \cite{mehlstaubler_atomic_2018,denker_geodetic_2018,delva_atomic_2013} such as offshore islands without direct line of sight to reference points on the mainland. Such remote frequency comparisons could in the future help to significantly refine existing geoid maps derived from satellite measurements. Moreover, a transportable clock could enable frequency comparisons among distant stationary optical clocks that are not connected via length-stabilized optical fibers, using sequential side-by-side frequency comparisons.

To reduce the total systematic fractional frequency uncertainty, data post-processing to exclude measurements affected by detected background gas collisions could be readily employed in our system. Furthermore, by replacing the multi-layer trap by one of the same design fabricated from a highly heat-conductive material such as AlN \cite{keller_spectroscopic_2015}, the uncertainty of the ions' thermal environment, and therefore the BBR shift, could be further suppressed. Moreover, AlN is much stiffer than the Rogers\textsuperscript{\textregistered} material employed here. Therefore, we expect it to bend less, resulting in a more accurate electrode geometry and a further reduced sensitivity of the ions' micromotion to axial positioning. The segmented trap structure offers several opportunities to improve the stability of the \alca clock. For instance, in a cascaded clock scheme the clock laser could be prestabilized to a large \ca Coulomb crystal \cite{aharon_robust_2018} using dynamical decoupling to suppress inhomogeneous systematic shifts across the crystal. This would enable longer \al clock ion interrogation times and thus improved statistical uncertainty \cite{peik_laser_2006,leroux_-line_2017}. Alternatively, a crystal consisting of multiple, possibly entangled \al and \ca ions, could be employed, ideally combined with a readout technique that only requires a logarithmic overhead in \ca ions \cite{schulte_quantum_2016} to minimize the axial extent of the crystal and therefore the EMM second-order Doppler shift as far as possible.

\begin{acknowledgments}
We thank M.~Dolezal and P.~Balling from CMI for performing the thermography measurements of rf heated trap chips and S.~A.~King for stimulating discussion. We acknowledge support from PTB and DFG through CRC 1227 (DQ-\textit{mat}), project B03 and CRC 1128 (geo-\textit{Q}), project A03. This project has received funding from the European Metrology Programme for Innovation and Research (EMPIR) co-financed by the participating states and from the European Union’s Horizon 2020 research and innovation programme, project No. 15SIB03 OC18. We acknowledge financial support by the Ministry of Science and Culture of Lower Saxony from “Nieders{\"a}chsisches Vorab” through “Fundamentals of Physics and Metrology (FPM)” initiative. 

\end{acknowledgments}

\section{Appendix}
\label{sec:app1}

\begin{figure}[htbp!]
\includegraphics[width=8.0cm]{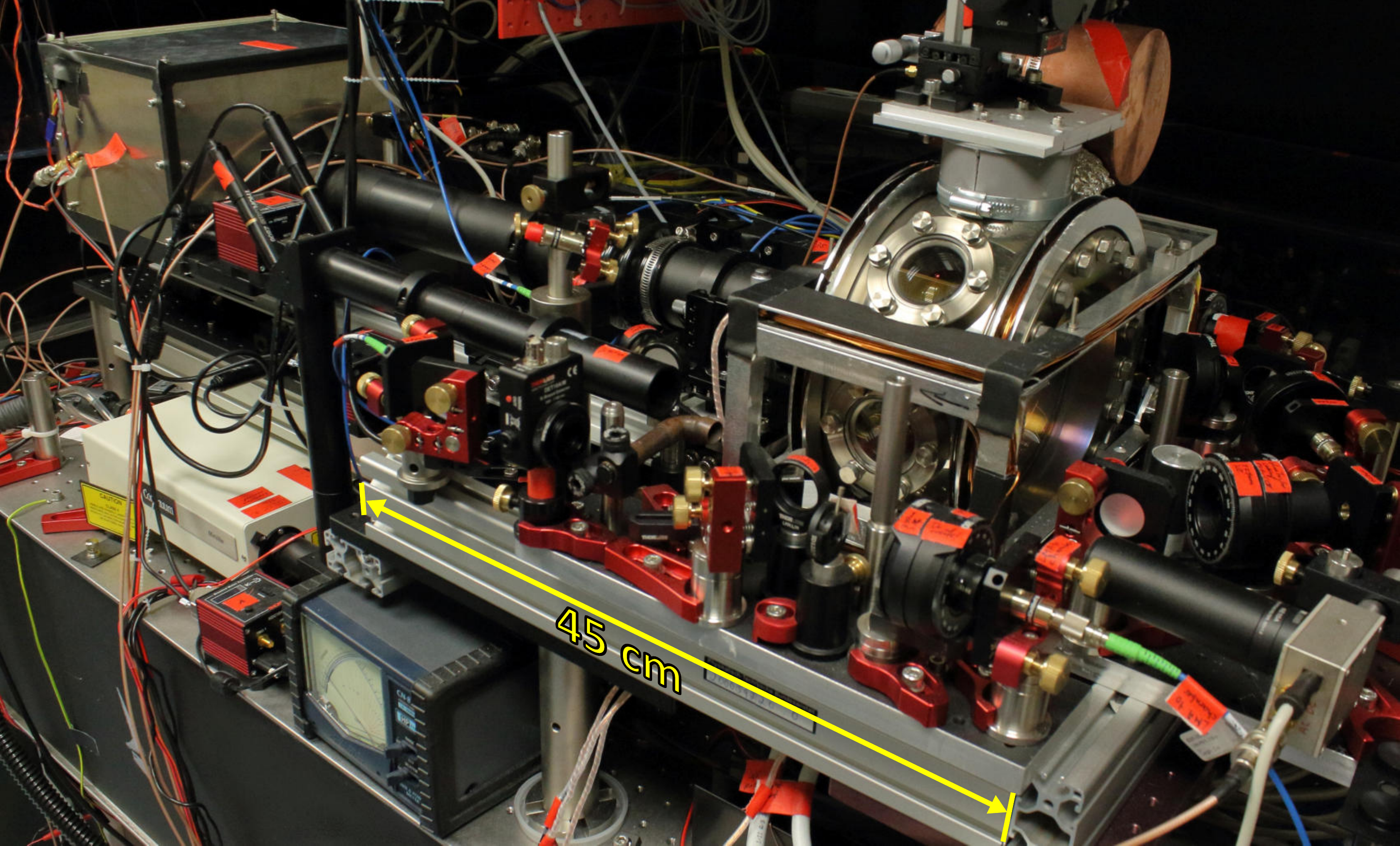}
\caption{Photo of the setup showing the vacuum vessel on the right and the imaging system on the left. The footprint of the system is $1.4\,\mathrm{m}\,\times\,0.5\,\mathrm{m}$.}
\label{fig:photo_setup}
\end{figure}

\begin{table} [htbp!]
\caption{Biaspherical lens surface profile parameters according to ISO 10110. The front surface points towards the ion trap, see Fig.~\ref{fig:achamber}.}
\label{Tab_lens}
\begin{small}
\begin{center}
\begin{ruledtabular}
\begin{tabular}{lll} 
Parameter & front surface & rear surface
\tabularnewline
\toprule
$R\ (\mathrm{mm})$ & $75.461975$ & $25.657609$
\tabularnewline
$k$ & $-13.754$ & $-0.88586$
\tabularnewline
$A_4\ (\mathrm{mm}^{-3})$ & $-1.8442\times10^{-6}$ & $1.8439\times10^{-6}$
\tabularnewline
$A_6\ (\mathrm{mm}^{-5})$ & $1.2836\times10^{-9}$ & $7.4104\times10^{-10}$
\tabularnewline
$A_8\ (\mathrm{mm}^{-7})$ & $1.2836\times10^{-9}$ & $2.7227\times10^{-13}$
\tabularnewline
\end{tabular}
\end{ruledtabular}
\end{center}
\end{small}
\end{table}




%

\end{document}